\documentclass[12pt,epsf]{article}
 \usepackage[dvips]{graphicx}
 \usepackage{verbatim,graphics,graphicx,color,slashed,amsmath}
 \usepackage{bm}
 \usepackage{cite}
 \usepackage{amssymb}
 \usepackage{braket}
 \usepackage{amsmath}
 \usepackage{bbding}
 \usepackage{xcolor}
  \usepackage{color}  
 
 \usepackage{hyperref}

\hypersetup{colorlinks,bookmarks,unicode,linktocpage=true,linkcolor=blue, anchorcolor=blue, citecolor=blue}

\newcounter{num}

\newcommand{\mll}{\ensuremath{m_{\ell\ell}}}
\newcommand{\hll}{\ensuremath{\eta_{\ell\ell}}}

\begin{document}
\thispagestyle{empty}
\vspace*{-15mm}
\begin{flushright}
{\footnotesize USTC-ICTS/PCFT-22-12}\\
{\footnotesize KCL-PH-TH/2022-09}
\end{flushright}
\vspace{15mm}
\begin{center}
{\Large\bf
Moments for positivity: using Drell-Yan data to test positivity bounds and reverse-engineer new physics
}
\vspace{7mm}

\baselineskip 18pt
{\bf Xu Li${}^{1,2,\, *}$, Ken Mimasu${}^{3,\, \ddagger}$, Kimiko Yamashita${}^{1,4,\, \S}$, \\
Chengjie Yang${}^{1,2,\, \P}$, Cen Zhang${}^{1,2,5,\, \dagger}$, Shuang-Yong Zhou${}^{6,7,\, \parallel}$}
\vspace{2mm}

{\it
${}^{1}$Institute of High Energy Physics, Chinese Academy of Sciences, Beijing 100049, China\\
${}^{2}$School of Physical Sciences, University of Chinese Academy of Sciences, Beijing 100049, China\\
${}^{3}$Theoretical Particle Physics and Cosmology Group, Department of Physics, King's College London, London WC2R 2LS, UK\\
${}^{4}$Department of Physics, Chung-Ang University, Seoul 06974, Korea\\
${}^{5}$Center for High Energy Physics, Peking University, Beijing 100871, China\\
${}^{6}$Interdisciplinary Center for Theoretical Study, University of Science and Technology of China, Hefei, Anhui 230026, China\\
${}^{7}$Peng Huanwu Center for Fundamental Theory, Hefei, Anhui 230026, China
\newline \newline
${}^{*}$lixu96@ihep.ac.cn,
${}^{\ddagger}$ken.mimasu@kcl.ac.uk,
${}^{\S}$kimikoy@cau.ac.kr,
${}^{\P}$yangchengjie@ihep.ac.cn,
${}^{\dagger}$Deceased June 2021,
${}^{\parallel}$zhoushy@ustc.edu.cn}\\
\vspace{10mm}
\end{center}
\begin{center}
\begin{minipage}{14cm}
\baselineskip 16pt
\noindent
\begin{abstract}
Moments of the leptonic angular distribution in the Drell-Yan process have recently been shown to be sensitive probes of a specific class of dimension-8, four-fermion operators in the Standard Model Effective Field Theory, involving a pair of quarks and leptons. The same operators are also subject to positivity bounds, when requiring the associated (unknown) UV completion to obey basic principles of quantum field theory. We perform a phenomenological study to quantify the sensitivity of the high-luminosity LHC to this set of operators and, by extension, the positivity bounds. We further extend the angular basis of moments and consider double differential information to improve the ability to disentangle the different operators, leading to a sensitivity to new physics scales up to 3 TeV. We use this information to explore the violation of positivity at the LHC as a way to test the underlying principles of quantum field theory. Finally, we present a case study which combines our results with information from other (current and prospective) experiments, as well as the positivity cone to infer the properties of possible tree-level UV completions.  The data lead to robust, model-independent lower bounds on the $M/\sqrt{g}$ combination of the particle mass and coupling, for states that couple to right-handed leptons and/or up quarks.\\
~\\
~\\
\begin{center}
    {\it In memory of Cen Zhang}
\end{center}
\end{abstract}
\end{minipage}
\end{center}

\baselineskip 18pt

\newpage

\tableofcontents

\section{Introduction}\label{sec:introduction}

The last decade has seen the rise to prominence of indirect searches for new physics at the Large Hadron Collider (LHC). The need for new physics beyond the Standard Model (SM), and the compelling possibility that it resides at the TeV scale, is in tension with the lack of evidence for new states in the LHC data. It is then plausible that these states lie just out of reach, in terms of direct production, but may nevertheless lead to deviations from SM predictions at high-energies that could be detected by the increasingly precise measurements of the LHC experiments. The Standard Model Effective Field Theory (SMEFT) has become the established framework to interpret data in the context of such deviations. The theory supplements the SM with higher dimensional operators that respect its symmetries and field content, and are suppressed by powers of the cutoff parameter, $\Lambda$, representing a generic scale at which new physics is expected to reside. 

The leading, non-trivial order at which the SMEFT operator expansion is truncated is $\mathcal{O}(1/\Lambda^2)$, \emph{i.e.}, involving operators of mass dim-6. (Henceforth, when referring to effective operators, we will abbreviate ``dimension-$n$'' by ``dim-$n$''.) This class of operators yields contributions to physical observables at both $\mathcal{O}(1/\Lambda^2)$ and $\mathcal{O}(1/\Lambda^4)$. The $\mathcal{O}(1/\Lambda^2)$ contribution comes from the interference between the dim-6 operator and SM amplitudes and is referred to as the ``interference" or ``linear'' term, while the $\mathcal{O}(1/\Lambda^4)$ piece comes from the square of the dim-6 amplitude, and we refer to it as the ``quadratic'' or ``squared'' term. Although the latter are formally beyond the desired leading order of the SMEFT, there are a number of reasons for which they might be important. For instance, the leading term, which arises from an interference between SM and SMEFT amplitudes, may be suppressed due to symmetries or helicity selection rules~\cite{Azatov:2016sqh}. In practice, this means that our current sensitivity to certain sectors of SMEFT operators can depend on whether such $\mathcal{O}(1/\Lambda^4)$ terms are included or not. However, for a complete description of SMEFT effects at $\mathcal{O}(1/\Lambda^4)$, one should also consider linear (interference) contributions from dim-8 operators, which also come in at this order. Although a fully general analysis up to $\mathcal{O}(1/\Lambda^4)$ is often difficult, mainly due to the large number of dim-8 operators, it is clearly of interest to study them from both the theoretical and phenomenological perspective. 

One of the pillars of the LHC physics programme is to measure the parameters of the SMEFT, such that we can make statements about possible new physics at the multi-TeV scale and steer the direction of future higher energy physics experiments. To this end, it is crucial that the precision programme of the LHC experiments be continuously developed. Aside from improved measurement techniques and more precise predictions for the SM and beyond, new observables should be sought to complement existing searches. These can be especially powerful when they single out unique features of new physics signatures and/or exploit the known properties of SM contributions to processes of interest. In this work we investigate moments of the leptonic angular distribution in the Drell-Yan process as such a candidate observable. As we will discuss in detail later on, certain angular moments have been shown to be especially sensitive to a class of dim-8 operators that would be indistinguishable from other dim-8 operators in non-angular observables such as the usual di-lepton invariant mass distribution~\cite{Alioli:2020kez}. The moments in question extract  the coefficients of the partial wave of total angular momentum $l\geq 3$. At leading order both the SM and dim-6 operator contributions to the $q\bar{q}\to\ell^-\ell^+$ amplitude have $l\leq 1$, so they cannot affect the $l>2$ moments. This picture is unchanged by QCD corrections and the first SM contributions to $l\geq2$ arise from subleading EW logarithms. The $l\geq3$ partial waves are therefore clean probes of pure dim-8 effects in the Drell-Yan processes. 

One of the main motivations for gaining access to dim-8 coefficients is that it allows us, for the first time, to test if the fundamental principles of quantum field theory are obeyed in physics beyond the SM. This can be achieved by checking whether a particular set of Wilson coefficients of the SMEFT are consistent with so-called positivity bounds \cite{Fuks:2020ujk}. Positivity bounds are constraints on the Wilson coefficients that can be bootstrapped from assuming that the UV completion of the SMEFT satisfies the axiomatic principles such as unitarity, causality/analyticity and locality \cite{Adams:2006sv} (see also earlier works \cite{Pham:1985cr, Ananthanarayan:1994hf}; see \cite{deRham:2022hpx} for a recent review). That is, they are independent of the specific details of the UV completion and can be taken as a proxy of robustness of the fundamental principles of quantum field theory. Recent years have seen increased interest in the high energy physics community, in extending the scope and applicability of positivity bounds \cite{Zhang:2020jyn, Li:2021lpe, deRham:2017avq, deRham:2017zjm, Arkani-Hamed:2020blm, Bellazzini:2020cot, Tolley:2020gtv, Caron-Huot:2020cmc, Sinha:2020win, Alberte:2020jsk, Chiang:2021ziz, Caron-Huot:2021rmr, Grall:2021xxm, Bern:2021ppb, Alberte:2021dnj, Du:2021byy, Bellazzini:2021oaj, Chowdhury:2021ynh, Caron-Huot:2022ugt, Chiang:2022jep, Chiang:2022ltp, Haring:2022cyf}. These bounds imply that the naive parameter space of an EFT as furnished by the independent Wilson coefficients is actually severely constrained by the fact that the EFT must be UV completed at high energies. 

In constraining the SMEFT up to $\mathcal{O}(1/\Lambda^4)$ at tree-level, only the leading $s^2$ positivity bounds are phenomenologically relevant. Due to the large number of dim-8 operators in the SMEFT, it is already non-trivial to extract the full information from these leading bounds~\cite{Zhang:2020jyn, Li:2021lpe}. The optimal $s^2$ positivity bounds can be computed by viewing the twice subtracted amplitudes as forming a convex cone. In practice, the positivity cone can be constructed either by first identifying its extremal rays and then converting to the positivity bounds ({\it i.e.}, the boundaries of the positivity cone)~\cite{Zhang:2020jyn} (see also~\cite{Bellazzini:2014waa}) or by directly computing the extremal rays of its dual cone~\cite{Li:2021lpe}. The former method involves constructing the group projectors of irreducible representations from the Clebsch-Gordon coefficients and is very effective for a problem with sufficient symmetries and a manageable number of degrees of freedom, while the latter is useful for generic problems that lack sufficient symmetries and consequently, the bounds are only extracted numerically in general. It is found that only a small fraction of the parameter space of the SMEFT is consistent with the positivity bounds, as discussed in~\cite{Zhang:2018shp, Bi:2019phv, Yamashita:2020gtt, Zhang:2021eeo, Remmen:2019cyz, Trott:2020ebl} for the example of constraining the anomalous quartic gauge couplings from vector boson scatterings. In~\cite{Fuks:2020ujk}, it was found that future electron-positron colliders can be used to probe violations of positivity bounds or the fundamental principles of quantum field theory up to the multi-TeV scale, regardless of the presence of dim-6 operators.

The fact that the extremal rays of the positivity cone are the group projectors of irreducible representations
in the product decomposition of the symmetries of an EFT implies that positivity bounds can be used to infer the UV states \cite{Zhang:2020jyn}. This is most significant if experiments were to find some new physics located close to the boundary of the positivity cone, or even better, close to one of its extremal rays~\cite{Fuks:2020ujk, Zhang:2021eeo}. In these favorable circumstances, comparing the positivity cone with the experimental data would enable the model-independent inference of UV particles' quantum numbers. On the other hand, should the future data indicate no deviation from the SM from dim-8 operators, then the convex nature of the positivity cone would, in principle, allow us to conclude that no new physics is present up to the scale probed by the experiments. This is different from experimentally constraining the scales of dim-6 operators, which do not form a convex cone, {\it i.e.}, nor satisfy positivity bounds, as there can be cancellations between different UV states that produce null dim-6 coefficients. Therefore, excluding the dim-6 operators up to a certain scale does not necessarily mean that the SM is valid up to that scale and new physics has to be at a higher scale. This provides another motivation to study dim-8 operators, despite the fact that they are subleading compared to the dim-6 operators, as a way to reverse-engineer UV models from the EFT interpretation of the data. Other applications of positivity bounds in the SMEFT can be found in~\cite{Gu:2020ldn, Vecchi:2007na, Bellazzini:2018paj, Remmen:2020vts, Bonnefoy:2020yee, Remmen:2020uze, Chala:2021wpj,Bellazzini:2017bkb}. Finally, we note that positivity bounds have also been widely used to constrain the EFT parameter spaces in other areas of particle physics and cosmology, leading to various interesting results ({\it e.g.}, \cite{Distler:2006if, Manohar:2008tc, Bellazzini:2015cra, Bellazzini:2016xrt, Cheung:2016yqr, Bonifacio:2016wcb, deRham:2017imi, Bellazzini:2017fep, deRham:2018qqo, Bonifacio:2018vzv, Melville:2019wyy, deRham:2019ctd, Alberte:2019xfh, Herrero-Valea:2019hde, Chen:2019qvr, Alberte:2020bdz, Huang:2020nqy, Tokuda:2020mlf, Wang:2020xlt, Wang:2020jxr, Herrero-Valea:2020wxz, deRham:2021fpu, Traykova:2021hbr, Arkani-Hamed:2021ajd, Haldar:2021rri, Raman:2021pkf, Gopakumar:2021dvg, Zahed:2021ffy, Kundu:2021qpi, Davighi:2021osh, Davis:2021oce, Alvarez:2021kpq, Melville:2022ykg, Li:2022tcz, Henriksson:2022oeu, Albert:2022oes}).

The paper is organized as follows. In Section~\ref{sec:eft}, we review the decomposition of the leptonic angular distribution into spherical harmonics, introducing the observables associated to the $l\geq3$ moments, $B_0$  and $D_0$. We then enumerate the relevant effective operators, discussing how they contribute to the $l\geq 3$ moments and their connection to the positivity bounds via the forward, elastic scattering amplitude. In Sec.~\ref{sec:positive}, we present a pedagogical review of how positivity bounds on effective operators are obtained, and obtain the elastic positivity bounds on our operators of interest. The angular dependence of the operator contributions to the Drell-Yan process as well as the master formulae for calculating $B_0$  and $D_0$, are presented and analyzed in Sec.~\ref{sec:ang}. Sec.~\ref{sec:cutoff} documents our phenomenological analysis, in which we construct and analyse the likelihood, and obtain projected sensitivities to the Wilson coefficients for the High-luminosity LHC (HL-LHC). We also study the sensitivity to the scale of positivity violation, in the event that evidence for non-zero Wilson coefficients is observed in Sec.~\ref{sec:positivity_v}. Finally, in Sec.~\ref{sec:positivity_uv}, we present a case study in which we take one operator from our study and combine it with other, related operators to derive the cone of extremal positivity bounds. Using the cone in conjuction with current and projected data from our study and the existing literature, we determine the novel implications for relevant UV states which include more robust, model-independent bounds on the new physics scale. We summarise and conclude in~Section~\ref{sec:summary}.

\section{Effective operators}\label{sec:eft}

The SMEFT approach augments the Standard Model with all possible higher dimensional operators that are built out of the Standard Model particles and consistent with the Standard Model symmetries in the unbroken phase:
\begin{equation}
\mathcal{L}_{\mathrm{SMEFT}} =  \mathcal{L}_{\mathrm{SM}} 
+  \sum_i \frac{C^{(6)}_i}{\Lambda^2}O^{(6)}_i
+ \sum_i \frac{C^{(8)}_i}{\Lambda^4}O^{(8)}_i + \cdots,
\end{equation}
where $\mathcal{L}_{\mathrm{SM}}$ is the renormalizable SM Lagrangian, $\Lambda$ is the cutoff scale, and the $C^{(j)}_i$ parameters are the (dimensionless) Wilson coefficients associated with the higher-dimensional operators. We have, for our purposes, assumed lepton/baryon number conservation, which forbids dim-5 and dim-7 operators.
Needless to say, this is a Lagrangian with a large number of free parameters, even if we only concern ourselves with specific processes. However, as discussed in the introduction, these Wilson coefficients are not allowed to take arbitrary values, assuming that the underlying UV theory that generates them obeys some basic principles of QFT. 

In this paper, we study the Drell-Yan process, $pp\to\ell^-\ell^+$, in proton-proton collisions. The kinematics of the process is often parametrised by  the invariant mass and rapidity of the dilepton system,  $\mll$ and $\hll$, and the polar and azimuthal angles of one of the leptons in the di-lepton rest frame, denoted by $\theta$ and $\phi$, respectively. Most often, collider experiments measure the $m_{\ell\ell}$ spectrum and the differential angular distributions of this final state. The angular dependence can be factorised and described as an expansion in spherical harmonics up to arbitrarily high total angular momentum. Following Ref.~\cite{Alioli:2020kez}, the expansion up to $l=3$ is given by:
\begin{equation}
\begin{aligned}
\frac{d \sigma_{pp \rightarrow \ell^- \ell^+}}{d \mll d \hll d \Omega_{\ell}}=& \frac{3}{16 \pi} \frac{d \sigma_{pp \rightarrow \ell^- \ell^+}}{d \mll d \hll}\left[\left(1+c_{\theta}^{2}\right)+\frac{\tilde{A}_{0}}{2}\left(1-3 c_{\theta}^{2}\right) +\tilde{A}_{1} s_{2 \theta} c_{\phi}+\frac{\tilde{A}_{2}}{2} s_{\theta}^{2} c_{2 \phi}+\tilde{A}_{3} s_{\theta} c_{\phi}\right. \\ &+\tilde{A}_{4} c_{\theta} 
+\tilde{A}_{5} s_{\theta}^{2} s_{2 \phi}+\tilde{A}_{6} s_{2 \theta} s_{\phi}+\tilde{A}_{7} s_{\theta} s_{\phi} 
+\tilde{B}_{3}^{e} s_{\theta}^{3} c_{3\phi}+\tilde{B}_{3}^{o} s_{\theta}^{3} s_{3\phi}+\tilde{B}_{2}^{e} s_{\theta}^{2} c_{\theta} c_{2 \phi} \\
&\left.+\tilde{B}_{2}^{o} s_{\theta}^{2} c_{\theta} s_{2 \phi}+\frac{\tilde{B}_{1}^{e}}{2} s_{\theta}\left(5 c_{\theta}^{2}-1\right) c_{\phi} +\frac{\tilde{B}_{1}^{o}}{2} s_{\theta}\left(5 c_{\theta}^{2}-1\right) s_{\phi}+\frac{\tilde{B}_{0}}{2}\left(5 c_{\theta}^{3}-3 c_{\theta}\right)\right],
\label{eq:ang_para}
\end{aligned}
\end{equation}
where $c_{\alpha}$ and  $s_{\alpha}$ represent the sine and cosine of an angle $\alpha$ respectively, $\Omega_{l}$ is the solid angle of the negatively charged lepton and all lepton angles are defined in the Collins-Soper frame~\cite{Collins:1977iv}.
The Collins-Soper frame corresponds to the rest frame of the dilepton system, with the $z$-axis defined as the external bisector of the angle between the momenta of the two protons at the LHC~\cite{ATLAS:2016rnf}.
At Leading Order (LO), the transverse momentum of the dilepton system is zero, and  the $z$-axis is aligned with the proton beam.
The positive $z$-axis is chosen based on the longitudinal direction of the dilepton system in the laboratory frame. Finally the $x$-axis points along the positive transverse momentum  direction of the dilepton system.
When there is no partonic transverse momentum, the direction of the $x$-axis and the definition of the azimuthal angle is arbitrary. Our SMEFT predictions are all taken at LO, while for the Next-to-Leading Order (NLO) QCD predictions from the SM, we integrate over the partonic transverse momentum.

The coefficients in Eq.~\eqref{eq:ang_para} can be extracted by taking angular moments of spherical harmonic functions, $f(\theta,\phi)$, over the differential angular distribution. Theoretically, this is defined by the following:
\begin{align}
\label{eq:angularmoment}
\langle f(\theta,\phi) \rangle \equiv
\frac{
\int {d \Omega_{\ell} \ 
\frac{d \sigma_{pp \rightarrow \ell^- \ell^+}}
{d \mll d \hll d \Omega_{\ell}} 
\cdot f(\theta,\phi)}
}
{\frac{d \sigma_{pp \rightarrow \ell^- \ell^+}}{d \mll d \hll}
},
\end{align}
and amounts to taking a weighted average over an event sample\footnote{In practice, it is not possible to extract the coefficients in this exact way at collider experiments. This is because of the fact that the finite detector acceptance limits the maximum scattering angle that can be observed, introducing an effective cut in $c_{\theta}$ that spoils the orthogonality of the spherical harmonics. Typically, analyses resort to constructing templates that account for the effect of finite angular acceptance on the angular dependencies associated to each coefficient and fitting the data to these template functions to extract them (see e.g.~\cite{CMS:2015cyj,ATLAS:2016rnf}).}. 
The $\tilde{A}_i$ coefficients are generated by the SM process and its QCD corrections, and have been measured by the LHC experiments~\cite{ATLAS:2012au,CMS:2011kaj,CMS:2015cyj,ATLAS:2016rnf,LHCb:2022tbc}. They are associated to the spherical harmonics of $l\leq 2$. The angular dependence associated to the $\tilde{B}_i$ coefficients, introduced in Ref.~\cite{Alioli:2020kez}, are described by combinations of the $l=3$ spherical harmonics, i.e.,
$Y^{\pm 3}_3$, $Y^{\pm 2}_3$, $Y^{\pm 1}_3$, and $Y^0_3 \equiv \sqrt{\frac{7}{16\pi}}(5 c_{\theta}^{3}-3 c_{\theta})$. The latter, associated to the coefficient $\tilde{B}_0$, will be the focus of our study, as it corresponds to the leading angular dependence arising from a particular class of dim-8 operators.  We also introduce the $\tilde{D}_i$ coefficients, which correspond to the angular moment of $l=4$ spherical harmonics, i.e., $Y^{\pm 4}_4, Y^{\pm 3}_4, Y^{\pm 2}_4, Y^{\pm 1}_4$ and $Y^{0}_4$ = $\frac{3}{16} \sqrt{\frac{1}{\pi}}(35c_{\theta}^4 - 30c_{\theta}^2 + 3)$, since these will also be populated by our operators of interest and help us distinguish among them.
\begin{align}
    \frac{d \sigma_{pp \rightarrow \ell^- \ell^+}}{d \mll d \hll d \Omega_{\ell}}=& \frac{3}{16 \pi} \frac{d \sigma_{pp \rightarrow \ell^- \ell^+}}{d \mll d \hll}\left\{
    \cdots + \tilde{D}_{4}^{e} s^4_{\theta} c_{4\phi} + \tilde{D}_{4}^{o} s^4_{\theta} s_{4\phi} + \tilde{D}_{3}^{e} s^3_{\theta} c_{\theta} c_{3\phi}+ \tilde{D}_{3}^{o} s^3_{\theta} c_{\theta} s_{3\phi} \right. \nonumber \\
    &  +\tilde{D}_{2}^{e} s^2_{\theta}(7c^2_{\theta}-1) c_{2\phi} + \tilde{D}_{2}^{o} s^2_{\theta}(7c^2_{\theta}-1) s_{2\phi} + \tilde{D}_{1}^{e} s_{\theta} (7c^3_{\theta}-3c_{\theta}) c_{\phi} \nonumber \\
    & \left. + \tilde{D}_{1}^{o} s_{\theta} (7c^3_{\theta}-3c_{\theta}) s_{\phi} + \frac{\tilde{D}_{0}}{2} (35c^4_{\theta}-30c^2_{\theta}+3)
    \right\}
\end{align}
We note that an analogous method of moments has previously been considered in the context of measuring dim-6 SMEFT contributions in Electroweak Higgs production and decay~\cite{Banerjee:2019twi,Banerjee:2020vtm}.

There are several operators that can contribute to the production of lepton pairs at the LHC, but only the following subset affects the  $l \geq 3 $ spherical harmonics in the angular expansion when interfering with the SM amplitude~\cite{Alioli:2020kez}:
\begin{align}
O_{8, lq\partial 3} &= (\bar{\ell}\gamma_{\mu} \overleftrightarrow{D}_{\nu}\ell) (\bar{q}\gamma^{\mu} \overleftrightarrow{D}^{\nu}q) \label{eq:op1}\\
O_{8, lq\partial 4} &= (\bar{\ell}\tau^I\gamma_{\mu} \overleftrightarrow{D}_{\nu}\ell) (\bar{q}\tau^I\gamma^{\mu} \overleftrightarrow{D}^{\nu}q)\label{eq:op2}\\
O_{8, ed\partial 2} &= (\bar{e}\gamma_{\mu} \overleftrightarrow{D}_{\nu}e) (\bar{d}\gamma^{\mu} \overleftrightarrow{D}^{\nu}d) \label{eq:op3}\\
O_{8, eu\partial 2} &= (\bar{e}\gamma_{\mu} \overleftrightarrow{D}_{\nu}e) (\bar{u}\gamma^{\mu} \overleftrightarrow{D}^{\nu}u) \label{eq:op4}\\
O_{8, ld\partial 2} &= (\bar{\ell}\gamma_{\mu} \overleftrightarrow{D}_{\nu}\ell) (\bar{d}\gamma^{\mu} \overleftrightarrow{D}^{\nu}d) \label{eq:op5}\\
O_{8, lu\partial 2} &= (\bar{\ell}\gamma_{\mu} \overleftrightarrow{D}_{\nu}\ell) (\bar{u}\gamma^{\mu} \overleftrightarrow{D}^{\nu}u) \label{eq:op6}\\
O_{8, qe\partial 2} &= (\bar{e}\gamma_{\mu} \overleftrightarrow{D}_{\nu}e) (\bar{q}\gamma^{\mu} \overleftrightarrow{D}^{\nu}q)\label{eq:op7}
\end{align}
where $\ell$ and $q$ denote the left-handed lepton and quark SU(2) doublet respectively, $e$ and $u$ denote the right-handed lepton and up-quark singlet respectively, and $\tau^I$ is the $SU(2)$ Pauli matrix. 
We shall denote the dimensionless Wilson coefficients for the operators in Eqs.~\eqref{eq:op1}--\eqref{eq:op7} as
\begin{align}
\vec{C}^{(8)} &= (C_{8, lq\partial 3},C_{8, lq\partial 4},C_{8, ed\partial 2}
,C_{8, eu\partial 2},C_{8, ld\partial 2},C_{8, lu\partial 2},C_{8, qe\partial 2}).
\end{align}
We consider a flavor universal scenario in our study, taking a single coefficient for each operator to weight all diagonal flavor combinations of quarks and leptons, respectively.
The particular derivative structure of these operators yields an amplitude that depends on $t\sim s(c_\theta-1)/2$ and $t^2$. The SM amplitude, on the other hand, is mediated by spin-1 states and therefore cannot exceed a single power of $t$. This is not affected by QCD corrections which factorise. Interfering the two amplitudes leads to the minimum required angular dependence $\propto c_{\theta}^3$ to populate the $l\ge3$ angular moments. The first SM contributions to $l\ge3$ arise from subleading, Electroweak corrections involving box diagrams that generate next-to-leading logarithmic (NLL) Sudakov logarithms $\propto\log(t/m_W^2)$~\cite{Alioli:2020kez,Denner:2006jr}. 

The angular dependence of the operators in Eqs.~\eqref{eq:op1}--\eqref{eq:op7} can be contrasted with the other classes of dim-8 operators that mediate the same scattering amplitudes. Firstly, it is clear that two derivatives are required to induce the $t^2$ terms, which rules out the majority of operators that can modify $q \bar{q}\to\ell^-\ell^+$ at $\mathcal{O}(1/\Lambda^4)$ (See~\cite{Boughezal:2021tih} for a complete list). The only remaining class of operators has the schematic structure:
\begin{align}
    \label{eq:other_operators}
    \left(\bar{q}\gamma^\mu q\right)\partial^2 
    \left(\bar{\ell}\gamma_\mu \ell\right),
\end{align}
whose amplitude contributions do not depend on $t$, and therefore also do not induce the required angular dependence to populate the $B_i$.

 The $q\bar{q}\to\ell^-\ell^+$ amplitude squared relevant for the $l\ge3$ partial waves is then
\begin{align}
\label{eq:ampsq}
|\mathcal{M}|^2 &= |\mathcal{M}_{\mathrm{SM\, NLL\, EW}}|^2 + \sum_{i} \frac{C_i^{(8)}}{\Lambda^4}\Delta\mathcal{M}^{(8)}_i + \sum_{i\geq j}\frac{C_{i}^{(8)}C_{j}^{(8)}}{\Lambda^8}|\mathcal{M}^{(8)}_{ij}|^2,
\end{align}
where $\Delta\mathcal{M}^{(8)}_i$ is shorthand for the interference between the dim-8 and the SM amplitudes, and $C_i^{(8)}$ runs through all the components of the $\vec{C}^{(8)}$ vector defined above.
Although the leading contribution to the $l=3$ partial waves from each dim-8 operator arises at $\mathcal{O}(1/\Lambda^4)$, we also investigate the effect of quadratic terms $\propto 1/\Lambda^8$ which can also populate the same (and higher) moments. We note that dim-6 operators of the SMEFT do not lead to the new $l=3$ spherical harmonics contributions at leading order since they, like the SM in this approximation, only give $l=1$ contributions to the amplitude. The interference between dim-6 and dim-8 operators $\propto 1/\Lambda^6$ can contribute to the $l=3$ ($\tilde{B}_i$) but not the $l=4$ ($\tilde{D}_i$) harmonics. In our analysis, we take the simplifying assumption that the dim-6 operator coefficients are sufficiently constrained by LHC and other data to not contribute significantly to our observables. 

In summary, measuring the $l\geq3$ partial wave contributions to the differential Drell-Yan cross section via the $\tilde{B}_i, \tilde{D}_i$ coefficients affords direct access to a specific class of dim-8 operators. The leading SM and dim-6 effects, as well as the large set of other dim-8 operators that can mediate the $q\bar{q}\to\ell^-\ell^+$ scattering amplitude, do not yield the required angular dependence up to subleading EW Sudakov logarithms, and are therefore strongly suppressed in these observables. In our work, we focus on $\tilde{B}_0$ and $\tilde{D}_0$, which are the only coefficients that remain after azimuthal integration. 

Interestingly, the specific kinematic dependence that makes our operator set interesting for the higher partial waves in Drell-Yan ($q\bar{q}\to\ell^-\ell^+$ amplitudes $\propto t^2$) is, by crossing symmetry, precisely the kinematic dependence that allows for the derivation of positivity bounds on the coefficients, $\vec{C}^{(8)}$, from the elastic $q\ell\to q\ell$ scattering amplitude in the forward limit. These observables can therefore provide an experimental test of positivity, and consequently a window into the fundamental principles of QFT. In the next Section, we will review the origin and derivation of positivity bounds in more detail, and discuss how to go beyond elastic positivity via extremal rays.

\section{Positivity bounds}\label{sec:positive}
\subsection{Pedagogical review}\label{sec:positivityreview}
As mentioned in the introduction, positivity bounds are very reliable theoretical constraints that can be imposed on the Wilson coefficients, or more generally, physical quantities involving the associated scattering amplitudes. They can be deduced by merely assuming that the EFT has a UV completion that satisfies the principles of quantum field theory/S-matrix such as Lorentz invariance, unitarity, causality and locality. Unitarity is essentially the conservation of probability in quantum evolution for the UV theory, rather than the EFT. In particle scatterings, it allows us to obtain the generalized optical theorem, which states that the UV amplitude ${\cal M}_{ij\to kl}$ for the scattering $i+j\to k+l$ satisfies
\begin{equation}
\label{got1}
  \frac{1}{2i} \left( {\cal M}_{ij\to kl} -{\cal M}^*_{kl\to ij} \right) = \frac12 \sum_X\int d\Pi_X 
    {\cal M}_{ij\to X}{\cal M}^*_{kl\to X}
\end{equation}
where $X$ enumerates all possible intermediate states and $d\Pi_X$ is the Lorentz-invariant phase space measure. This contains information on positivity, the simplest of which can be extracted by the usual optical theorem in the forward limit:
\begin{equation}
        {\rm Im} { \cal M}_{ij\to ij} = \frac12 \sum_X\int d\Pi_X 
    |{\cal M}_{ij\to X}|^2 >0.
\end{equation}
This is the foundation for the simplest elastic postivity bounds. However, as pointed out in \cite{Zhang:2020jyn, Li:2021lpe}, the generalized optical theorem actually contains more positivity information, as the right hand side is a positive sum of ${\cal M}_{ij\to X}{\cal M}^*_{kl\to X}$. A positive sum is also known as a conical hull, so the generalized optical theorem defines an (abstract) convex cone structure, which is the engine for the convex cone (or extremal) positivity bounds \cite{Zhang:2020jyn, Li:2021lpe}. 

\begin{figure}[h!]
	\centering
		\includegraphics[width=.5\linewidth]{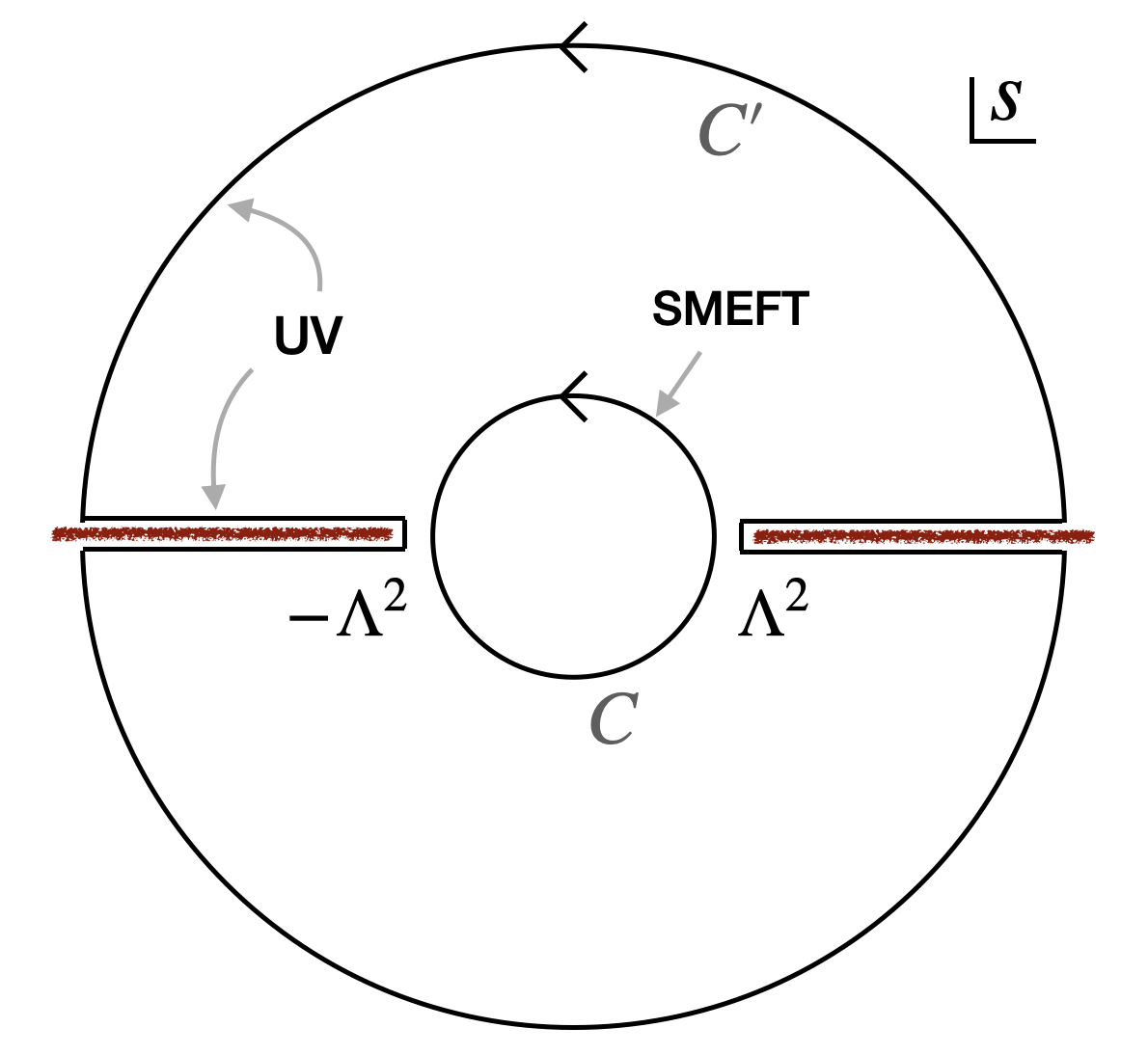}
	\caption{Contour integrals in the complex $s$ plane for the dispersion relation for ${d^2 M_{ijkl}(s)}/(2{d s^2})$. The small contour denoted by ``SMEFT'' can be evaluated at low energies within the SMEFT, while the big contour denoted by ``UV'', which can be deformed from the small contour, can only be fully evaluated in the UV completion. Even though the UV theory is unknown, the fundamental S-matrix properties of the UV theory provide useful information about this big contour $C'$, which, via the dispersion relation, can be used to constrain the small contour $C$ that contains information of the Wilson coefficients.}
	\label{fig:contour}
\end{figure}

Since we are concerned with tree-level EFT amplitudes including up to dim-8 operators, the highest order terms in energy growth are the $s^2$, $st$ or $t^2$ ones, once the poles in the amplitude are subtracted. On the other hand, as we shall see shortly, the positivity bounds come from the dispersion relation with two derivatives with respective to $s$, so we can focus on positivity bounds in the forward limit ($t=0$) without loss of information.  Causality means that one can view the UV amplitude ${\cal M}_{ij\to kl}(s,t=0)$ as an analytic function in the complex $s$ plane, with only poles and branch cuts on the real $s$ axis. For simplicity, we shall define a pole subtracted amplitude 
\begin{equation}
\label{defMijkl0}
M_{ijkl}(s)\equiv {\cal M}_{ij\to kl}(s,t=0)-({\rm low~energy~poles}),
\end{equation}
and then use Cauchy's integral formula to write the UV amplitude as a contour integral
\begin{equation}\label{disp1}
    \frac{1}{2} \frac{d^2 M_{ijkl}(s)}{d s^2} = \oint_{C} \frac{d \mu}{2\pi i} \frac{M_{ijkl}(\mu)}{(\mu-s)^3}
\end{equation}
where $C$ is a closed contour near $s=0$ avoiding the branch cuts (see Figure~\ref{fig:contour}). Due to the analyticity of the amplitude on the $s$ plane, we can deform the contour $C$ to infinity and to go around the branch cuts, obtaining the contour $C'$. This splits the right hand side of Eq.~(\ref{disp1}) into two parts: contributions from the two infinite semi-circles and contributions from the discontinuities of the branch cuts on the real $s$ axis. The contributions from the semi-circles actually vanish, due to the Froissart bound~\cite{Froissart:1961ux}, which states that any amplitude that satisfies unitarity and locality should grow slower than $s\ln^2{s}$ as $s\to \infty$ \footnote{Technically, the Froissart bound only works when the scattering particles are massive, but similar bounds are generally expected to be valid even for massless particles. At worst, locality only requires the amplitude to be polynomially bounded at large $s$.}. So we are left with
\begin{equation}
     \frac{1}{2} \frac{d^2 M_{ijkl}(s)}{d s^2} = \int^\infty_{-\infty} \frac{d \mu}{2\pi i} \frac{{\rm Disc} M_{ijkl}(\mu)}{(\mu-s)^3} 
\end{equation}
where the discontinuity is defined as ${\rm Disc}f(s)=f(s+i\epsilon)-f(s-i\epsilon)$. This is a twice subtracted dispersion relation, and furnishes a remarkable connection between the IR and the UV: when $s$ is small, the left hand side can be approximated by the EFT amplitude, representing the IR; on the other hand, we have the $\mu$ integration going up to infinity, so the right hand side represents unknown physics in the UV. As is standard, we shall assume that the SMEFT is weakly coupled below the cutoff $\Lambda$ so that we can approximate the SMEFT amplitude with the tree-level amplitude. Also, assuming that $\Lambda$ is much greater than the SM masses, we shall take the massless limit for the SM particles, to the leading approximation. Then, we can rewrite the dispersion relation, setting $s=0$, as
\begin{equation}
         \frac12 \frac{d^2 M_{ijkl}(0)}{d s^2} = \int^\infty_{\Lambda^2} \frac{d \mu}{2\pi i\mu^3} \left(
         {\rm{Disc}} M_{ijkl}(\mu) + {\rm Disc} M_{i\tilde lk\tilde j}(\mu)
         \right),
\end{equation}
where the integration below the energy scale of the lowest lying UV state $\Lambda^2$, which is identified with the EFT cutoff squared for simplicity, is dropped because the SMEFT tree-level amplitude with poles subtracted does not gives rise to any discontinuity, and we have also used crossing symmetry $M_{ijkl}(\mu)=M_{i\tilde lk\tilde j}(-\mu)$ in the massless, forward limit. Here, $\tilde{i}$ denotes the antiparticle of state $i$. Crossing symmetry is also a consequence of unitarity, causality and so on. Another basic property of the S-matrix is Hermitian analyticity: ${\cal M}^*_{kl\to ij}(s,t)={\cal M}_{ij\to kl}(s^*,t)$, by which the left hand side of Eq.~(\ref{got1}) becomes ${\rm Disc}{\cal M}_{ij\to kl}/(2i)$. Making use of the generalized optical theorem, the dispersion relation finally reduces to
\begin{equation}
\label{ampconvexC}
         \frac12 \frac{d^2 M_{ijkl}(0)}{d s^2} = \sum_X\int d\Pi_X\int^\infty_{\Lambda^2} \frac{d \mu}{2\pi \mu^3} \left( m_{ij}m^*_{kl} + m_{i\tilde l}m^*_{k\tilde j} \right),
\end{equation}
where we have defined $m_{ij}=M_{ij\to X}(\mu)$. If we take a model-independent approach for the UV theory, we should regard $m_{ij}$ as unknown complex numbers that depend on $\mu$, with $\mu$ essentially encoding the UV particle masses. Again, for an inelastic scattering, $m_{ij}m^*_{kl} + m_{i\tilde l}m^*_{k\tilde j}$ is generally not positive definite; the observation in \cite{Zhang:2020jyn, Li:2021lpe} is that the summation and the integration in front of $m_{ij}m^*_{kl} + m_{i\tilde l}m^*_{k\tilde j}$ is positive. This means that we can view ${d^2 M_{ijkl}(0)}/{d s^2}$ as a conical hull of the elements $m_{ij}m^*_{kl} + m_{i\tilde l}m^*_{k\tilde j}$. That is, the SMEFT amplitudes 
\begin{center}
{\it ${d^2 M_{ijkl}(0)}/{d s^2}$ with different $i,j,k,l$ form a convex cone,}
\end{center}
which constrains the Wilson coefficients to a conical subspace of the total parameter space. The convex cone positivity bounds generalize the elastic positivity bounds, which are just the special cases where the out-going $k$ and $l$ particle are chosen to be the same as the in-coming $i$ and $j$ particle:
\begin{equation}
\label{elasticbound}
 \frac12 \frac{d^2 M_{ijij}(0)}{d s^2}\geq 0
\end{equation}
We shall make use of both the elastic positivity bounds and the convex cone bounds in this paper. We emphasize that the positivity bounds are not a direct use of the generalized optical theorem, which is valid for the UV theory; rather, the positivity bounds exist because one can ``bring down'' the positivity of the UV theory to the EFT via the dispersion relation, which encodes analyticity and unitarity of the S-matrix. 

\subsection{Elastic positivity bounds on $qq\ell\ell$ operators}\label{sec:elasticpositivity}
For the operators in Eqs.~(\ref{eq:op1}-\ref{eq:op7}), we shall primarily consider the elastic bounds. That is, we compute the tree-level elastic amplitudes for leptons and quarks with different species and chiralities and impose the inequality \eqref{elasticbound}. In the forward limit, only the 7 operators in question give non-zero contributions to the twice-subtracted dispersion relation. One can see that the other, similar operators with schematic form given in Eq.~\eqref{eq:other_operators} do not contribute to the elastic process in the forward limit since, by crossing symmetry, they only contribute like $t^2$. The non-trivial bounds are summarized in Table~\ref{tab:elastic_bounds}. The bounds listed are directly results of Eq.~\eqref{elasticbound}, retaining the factors of $-4$, which will be useful for later purposes. Generally, each operator contributes to a different species/helicity configuration, leading to a set of nearly independent elastic positivity constraints. Only the $\mathcal{O}_{8,ql\partial 3}$ and $\mathcal{O}_{8,ql\partial 4}$ operators are correlated by elastic positivity as they both contribute to the fully left-handed $d\ell\to d\ell$ and $u\ell\to u\ell$ amplitudes. 

The simplicity and independence of the bounds on other operators beyond those of Eqs.~(\ref{eq:op1}-\ref{eq:op7}) is the main reason why we restrict ourselves to the elastic positivity bounds. This allows us to focus on the connection between these bounds and the $B_0$ angular moments in the Drell-Yan process.
Conversely, the convex cone positivity bounds would involve many more operators, which makes their derivation a formidable task that we leave for future work. Since the extra operators entering the convex cone positivity bounds would be unconstrained by the angular moments considered in this paper, it is {\it a priori} unclear how much we would benefit from imposing them, in terms of constraining the Wilson coefficients space and therefore testing positivity. A complete understanding of picture would require a global analysis of a larger set of data and dimension-8 operators that is beyond the scope of this work.

\begin{table}[ht]
\centering
\begin{tabular}{c|c}
\hline 
Positivity bound & channel: $\ket{1} +\ket{2} \to   \ket{1} +\ket{2}$\\
\hline \hline
$-4C_{8, lq\partial 3}+4C_{8, lq\partial 4} \geq 0$ & $\ket{1}=\ket{e^{-}_L},~\ket{2}=\ket{u_L}$\\
$-4C_{8, lq\partial 3}-4C_{8, lq\partial 4} \geq 0$ & $\ket{1}=\ket{e^{-}_L},~\ket{2}=\ket{d_L}$\\
$-4C_{8, ed\partial 2}	\geq 0$ & $\ket{1}=\ket{e^{-}_R},~\ket{2}=\ket{d_R}$\\
$-4C_{8, eu\partial 2}	\geq 0$ & $\ket{1}=\ket{e^{-}_R},~\ket{2}=\ket{u_R}$\\
$-4C_{8, ld\partial 2}	\geq 0$ & $\ket{1}=\ket{e^{-}_L},~\ket{2}=\ket{d_R}$\\
$-4C_{8, lu\partial 2}	\geq 0$ & $\ket{1}=\ket{e^{-}_L},~\ket{2}=\ket{u_R}$\\
$-4C_{8, qe\partial 2}	\geq 0$ & $\ket{1}=\ket{e^{-}_R},~\ket{2}=\ket{u_L}$\\
\hline
\end{tabular}
\caption{Elastic positivity bounds for the operators (\ref{eq:op1}-\ref{eq:op7}).  The right column shows the scattering channel for the corresponding elastic bound.}
\label{tab:elastic_bounds}
\end{table}
\subsection{Positivity cone and the inverse problem}

One important feature of the convex cone bounds is that the extremal rays (ER) of the positivity cone can often have important physical significance \cite{Zhang:2020jyn}. This is especially true when the EFT is constrained by symmetries such as in the case of the SMEFT. An ER is an element of the convex cone that cannot be represented by a sum of two other elements. Since the convex cone of the amplitude ${d^2 M_{ijkl}(0)}/{d s^2}$ is obtained as a conical hull of $m_{ij}m^*_{kl} + m_{i\tilde l}m^*_{k\tilde j}$, ERs must be of the form of $m_{ij}m^*_{kl} + m_{i\tilde l}m^*_{k\tilde j}$. Furthermore, since the $i,j,k,l$ particles live in some Lorentz and internal symmetry multiplets, the ERs correspond to the $m_{ij}$'s that are irreducible representations in the product decomposition of the representation of $i$ and $j$~\cite{Zhang:2020jyn, Fuks:2020ujk, Yamashita:2020gtt, Zhang:2021eeo}. When the symmetries of the theory are sufficiently strong for a particular problem, this approach with ERs from irreducible representations can constrain the convex cone to be mostly polyhedral, and we will mostly have isolated ERs. In the case of a tree-level UV completion, each of these ERs represents a UV particle with specific quantum numbers or symmetry structure that is a projection of the symmetry of UV theory to that of the SM~\cite{Zhang:2020jyn}. (For a loop-level UV completion, an ER may not directly correspond to the UV particle but becomes a component of the UV loop~\cite{Zhang:2021eeo}.) This implies that the convex cone bounds, or the dim-8 operators, are valuable and pertinent to the inverse-problem of reverse-engineering the UV model from a set of measured Wilson coefficients. 

In Section \ref{sec:positivity_uv}, we shall consider the simplified problem of obtaining the convex cone bounds for UV physics coupled to the right handed electron and up quarks, and show how to use them to infer the properties of all possible UV states. This is a simplified problem, because it only includes the $O_{8, eu\partial 2}$ operator, as opposed to all the 7 operators in, Eqs.~(\ref{eq:op1}-\ref{eq:op7}), that can be probed by the $B_0$ and $D_0$ quantities. Our projected sensitivity on its operator coefficient will help determine the degree to which the properties of all possible UV states that couple to right handed electrons and up quarks can be inferred. 
However, as mentioned before, the convex cone bounds will bring in other operators involving these states, which will require us to additionally make use of bounds derived from other processes and observables.

\section{Angular dependence}\label{sec:ang}
We now turn to the phenomenological analysis of the dim-8 contributions to Drell-Yan and the use of angular moments to constrain the operators introduced in Sec.~\ref{sec:eft}, Eqs.~\eqref{eq:op1}--\eqref{eq:op7}.
We start by obtaining the LO matrix elements for the partonic scattering process $q(p_1) \bar{q}(p_2) \to \ell^-(p_3) \ell^+(p_4)$. Here, $q$ $(\bar{q})$ denotes a quark (anti-quark) and $\ell^-$ $(\ell^+)$ denotes an electron or a muon (positron or anti-muon). We do not consider $\tau$ final states in our study and take all external states to be massless. 
The matrix-element squared of the SM contribution to each helicity configuration, denoted by $\{L,R\}$, is as follows:
\begin{align}
&|\mathcal{M}_\mathrm{SM}(q_L \bar{q}_R \to e^-_R e^+_L)|^2 =16\pi^2 \alpha^2(1-\hat{c}_\theta)^2\cdot
\left|Q_f-\frac{(I_3-s^2_WQ_f)\hat{s}}{c^2_W(\hat{s}-M^2_Z)}\right|^2, \label{eq:SM_M_squared}\\
&|\mathcal{M}_\mathrm{SM}(q_R \bar{q}_L \to e^-_R e^+_L)|^2  =16\pi^2 \alpha^2(1+\hat{c}_\theta)^2\cdot
\left|Q_f + \frac{s^2_W Q_f \hat{s}}{c^2_W(\hat{s}-M^2_Z)}\right|^2,\\
&|\mathcal{M}_\mathrm{SM}(q_L \bar{q}_R \to e^-_L e^+_R )|^2  =16\pi^2 \alpha^2(1+\hat{c}_\theta)^2\cdot
\left|Q_f -\frac{(I_3-s^2_WQ_f)(-1/2+s^2_W)\hat{s}}{c^2_Ws^2_W(\hat{s}-M^2_Z)}\right|^2,\\
&|\mathcal{M}_\mathrm{SM}(q_R \bar{q}_L \to e^-_L e^+_R)|^2  =16\pi^2 \alpha^2(1-\hat{c}_\theta)^2\cdot
\left|Q_f + \frac{Q_f(-1/2+s^2_W)\hat{s}}{c^2_W(\hat{s}-M^2_Z)}\right|^2,
\end{align}
where $\hat{s}$ is the partonic Mandelstam invariant $\hat{s} = (p_1 + p_2)^2 = m_{\ell \ell}^2$, $\hat{c}_{\theta}$ is the cosine of the polar angle between the incoming quark and the outgoing negatively charged lepton direction in the partonic center-of-mass frame. $\alpha$ is the fine-structure constant, $Q_f$ and $I_3$ are the electric charge and the third component of the quark isospin respectively, $c_W = \cos{\theta_W}$ and $s_W = \sin{\theta_W}$ with the Weinberg angle $\theta_W$, and $M_Z$ is the $Z$ boson mass. Hereafter, we neglect the $Z$ boson decay width as we will concentrate on the high-energy region $\sqrt{\hat{s}}= m_{\ell \ell} \geq 100$~GeV in our analysis. 

We find the leading contribution from each operator, corresponding to the interference with the SM amplitude (c.f. Eq.~\eqref{eq:ampsq}) to be
\begin{align}
\label{eq:deltaM_lq3}
&\Delta \mathcal{M}_{8, lq\partial 3} = -\frac{C_{8, lq\partial 3}}{\Lambda^4}8\pi\alpha \, \hat{c}_\theta(1+\hat{c}_\theta)^2 \hat{s}^2\cdot
\left(Q_f-\frac{(I_3-s^2_WQ_f)(-1/2+s^2_W)\hat{s}}{c^2_W s^2_W(\hat{s}-M^2_Z)}\right), \\
\label{eq:deltaM_lq4}
&\Delta \mathcal{M}_{8, lq\partial 4} = \frac{C_{8, lq\partial 4}}{\Lambda^4}8\pi\alpha \, \hat{c}_\theta(1+\hat{c}_\theta)^2 \hat{s}^2 (2I_3)\cdot
\left(Q_f-\frac{(I_3-s^2_WQ_f)(-1/2+s^2_W)\hat{s}}{c^2_W s^2_W(\hat{s}-M^2_Z)}\right), \\
\label{eq:deltaM_ed}
&\Delta\mathcal{M}_{8, ed\partial 2} = \frac{C_{8, ed\partial 2}}{\Lambda^4}8\pi\alpha \,\hat{c}_\theta(1+\hat{c}_\theta)^2 \hat{s}^2\cdot\frac{1}{3}
\left(1+\frac{s^2_W\hat{s}}{c^2_W(\hat{s}-M^2_Z)}\right), \\
\label{eq:deltaM_eu}
&\Delta\mathcal{M}_{8, eu\partial 2} = -\frac{C_{8, eu\partial 2}}{\Lambda^4}8\pi\alpha \,\hat{c}_\theta(1+\hat{c}_\theta)^2 \hat{s}^2\cdot\frac{2}{3}
\left(1+\frac{s^2_W\hat{s}}{c^2_W(\hat{s}-M^2_Z)}\right), \\
\label{eq:deltaM_ld}
&\Delta\mathcal{M}_{8, ld\partial 2} = \frac{C_{8, ld\partial 2}}{\Lambda^4}8\pi\alpha \,\hat{c}_\theta(1-\hat{c}_\theta)^2 \hat{s}^2\cdot\frac{1}{3}
\left(1+\frac{(-1/2+s^2_W)\hat{s}}{c^2_W(\hat{s}-M^2_Z)}\right), \\
\label{eq:deltaM_lu}
&\Delta\mathcal{M}_{8, lu\partial 2} = -\frac{C_{8, lu\partial 2}}{\Lambda^4}8\pi\alpha \,\hat{c}_\theta(1-\hat{c}_\theta)^2 \hat{s}^2\cdot\frac{2}{3}
\left(1+\frac{(-1/2+s^2_W)\hat{s}}{c^2_W(\hat{s}-M^2_Z)}\right) , \\
\label{eq:deltaM_qe}
&\Delta\mathcal{M}_{8, qe\partial 2} =-\frac{C_{8, qe\partial 2}}{\Lambda^4}8\pi\alpha \,\hat{c}_\theta(1-\hat{c}_\theta)^2 \hat{s}^2\cdot
\left(Q_f-\frac{(I_3-s^2_WQ_f)\hat{s}}{c^2_W(\hat{s}-M^2_Z)}\right).
\end{align}
The higher order angular dependence $\propto \hat{c}_{\theta}^3$ is present in all cases, as discussed in Sec.~\ref{sec:eft}.
Finally, the quadratic contribution to the matrix element squared from each operator is given by
\begin{align}
\label{eq:M8sq_lq3}
|\mathcal{M}_{8, lq\partial 3}|^2 &= \frac{C^2_{8, lq\partial 3}}{\Lambda^8}\hat{c}^2_\theta(1+\hat{c}_\theta)^2 \hat{s}^4, \\
\label{eq:M8sq_lq4}
|\mathcal{M}_{8, lq\partial 4}|^2 &= \frac{C^2_{8, lq\partial 4}}{\Lambda^8}\hat{c}^2_\theta(1+\hat{c}_\theta)^2 \hat{s}^4, \\
\label{eq:M8sq_ed}
|\mathcal{M}_{8, ed\partial 2}|^2 &= \frac{C^2_{8, ed\partial 2}}{\Lambda^8}\hat{c}^2_\theta(1+\hat{c}_\theta)^2 \hat{s}^4, \\
\label{eq:M8sq_eu}
|\mathcal{M}_{8, eu\partial 2}|^2 &= \frac{C^2_{8, eu\partial 2}}{\Lambda^8}\hat{c}^2_\theta(1+\hat{c}_\theta)^2 \hat{s}^4, \\
\label{eq:M8sq_ld}
|\mathcal{M}_{8, ld\partial 2}|^2 &= \frac{C^2_{8, ld\partial 2}}{\Lambda^8}\hat{c}^2_\theta(1-\hat{c}_\theta)^2 \hat{s}^4, \\
\label{eq:M8sq_lu}
|\mathcal{M}_{8, lu\partial 2}|^2 &= \frac{C^2_{8, lu\partial 2}}{\Lambda^8}\hat{c}^2_\theta(1-\hat{c}_\theta)^2 \hat{s}^4, \\
\label{eq:M8sq_qe}
|\mathcal{M}_{8, qe\partial 2}|^2 &= \frac{C^2_{8, qe\partial 2}}{\Lambda^8}\hat{c}^2_\theta(1-\hat{c}_\theta)^2 \hat{s}^4\\
|\mathcal{M}_{8, lq\partial 3,lq\partial 4}|^2 &= \frac{C_{8, lq\partial 3}C_{8,lq\partial 4}}{\Lambda^8}\hat{c}^2_\theta(1+\hat{c}_\theta)^2 \hat{s}^4. \label{eq:dim8_M_squared_end}
\end{align}
The last term corresponds to the interference between the amplitudes from $\mathcal{O}_{8, lq\partial 3}$ and $\mathcal{O}_{8, lq\partial 4}$.
Not only is the higher order angular dependence $\propto \hat{c}_{\theta}^3$ present, but a further dependence $\propto \hat{c}_{\theta}^4$ is also present in all cases.

The $pp\to \ell^- \ell^+$ cross section at the LHC is obtained by convoluting with the parton distribution functions (PDFs), and can be written:
\begin{align}
    \sigma_{p p \to \ell^- \ell^+} =\frac{1}{12} \sum_{ij} 
    \int\,dx_1\,dx_2\, G_{ij}(x_1,x_2,Q^2)
             \left[
             \int d\hat{c}_\theta
             \frac{d\hat{\sigma}}{d\hat{c}_\theta}(\hat{c}_\theta,\hat{s})
             \right], \label{eq:xsec}
\end{align}
 where the direction of parton 1 defines the positive $z$-axis,
\begin{align}
  G_{ij}(x_1,x_2,Q^2)=f_{i/p}(x_1,Q^2)\cdot f_{j/p}(x_2,Q^2)
\end{align} 
is the product of PDFs for the two possible partonic initial states, $ij=\{q\bar{q},\bar{q}q\}$ with $q=(u,d,c,s,b)$, 
and
$Q$ is PDF factorisation scale.  The partonic centre of mass energy, $\sqrt{\hat{s}}$, is equivalent to the invariant mass of leptons $m_{\ell \ell}$ at leading order and is related to the proton energy, $E_p$, by $\hat{s}=4 x_1 x_2 E_p^2$.
In the center-of-mass frame, taking the initial and final states to be massless, $\frac{d \sigma_{q \bar{q} \rightarrow \ell^- \ell^+}}{d \hat{c}_\theta}=\frac{1}{32\pi \hat{s}}|\mathcal{M}|^2$, with the different pieces of the matrix element squared given by Eqs.~\eqref{eq:SM_M_squared}--\eqref{eq:dim8_M_squared_end}, in which the initial spins are summed over, but the colors are not. Thus, the prefactor of $1/12=3/(3\cdot 3\cdot 4)$ in Eq.~\eqref{eq:xsec} arises from summing over the colors and taking the average of spins and colors.

Since the two partonic initial states have opposite $z$-axis orientations, the scattering angle $\hat{c}_\theta$ is not observable. We therefore convert $\hat{c}_\theta$ to $c_\theta^\ast$, the scattering angle in the centre-of-mass frame defined with respect to the positive $z$-axis in the lab frame, \emph{i.e.}, $\hat{c}_\theta=\pm c_\theta^\ast$ for the $q\bar{q}$ and $\bar{q}q$ initial states, respectively. We note that redefining an integration variable up to a sign has a trivial Jacobian when integrating over a symmetric interval. Changing also from
$dx_1\,dx_2\to d\tau\, d\eta = 2 (\sqrt{\hat{s}}/s) d\sqrt{\hat{s}}\, d\eta$ gives
\begin{align}
        \frac{d \sigma_{p p \to \ell^- \ell^+}}{d\sqrt{\hat{s}}\, d\eta\,dc_\theta^\ast\,}               =\frac{1}{12}\cdot \frac{2\sqrt{\hat{s}}}{s}
             \left[
             G_{q\bar{q}}(\hat{s},\eta,Q^2)
             \frac{d\hat{\sigma}}{d\hat{c}_\theta}(c_\theta^\ast,\hat{s})
             +
             G_{\bar{q}q}(\hat{s},\eta,Q^2)
             \frac{d\hat{\sigma}}{d\hat{c}_\theta}(-c_\theta^\ast,\hat{s})
             \right]
\end{align}
 with $\tau\equiv\hat{s}/s$, $s=4E^2_p$, and $\eta\equiv\frac{1}{2}\log(x_1/x_2)$. We finally want to take the scattering angle definition to be in the Collins-Soper frame, which (at leading order) orients the $z$-axis along the direction of $\eta$,
\begin{align}
    c_\theta=c_\theta^\ast\frac{\eta}{|\eta|}\equiv c_\theta^\ast\xi,
\end{align}
such that
\begin{align}
    \frac{d\sigma_{p p \to \ell^- \ell^+}}
    {d\mll\, d\hll\,dc_\theta}&=\frac{\mll}{6s}
             \left[
             G_{q\bar{q}}(\mll,\hll,Q^2)
             \frac{d\hat{\sigma}}{d\hat{c}_\theta}(c_\theta\xi,\mll)
             +
             G_{\bar{q}q}(\mll,\hll,Q^2)
             \frac{d\hat{\sigma}}{d\hat{c}_\theta}(-c_\theta\xi,\mll)
             \right],\label{eq:diffxsec}
\end{align}
where we now identify $\sqrt{\hat{s}}=\mll$ and $\eta=\hll$.
Integrating over the scattering angle gives:
\begin{align}
    \frac{d\sigma_{p p \to \ell^- \ell^+}}{d\mll\, d\hll}&= 
            \frac{\mll}{6s} \int_{-1}^{1}\,dc_\theta\,
              \left[
             G_{q\bar{q}}(\mll,\hll,Q^2)
             +
             G_{\bar{q}q}(\mll,\hll,Q^2)
             \right]
             \frac{d\hat{\sigma}}{d\hat{c}_\theta}(c_\theta\xi,\mll),
\end{align}
where we changed variable from $c_\theta\to -c_\theta$ in the second term. Similarly, the angular moment of an arbitrary function $f(c_{\theta})$ is obtained from Eq.~\eqref{eq:diffxsec} as follows. For even functions  of $c_{\theta}$, i.e., $f_{e}(-c_{\theta}) = f_{e}(c_{\theta})$, the angular moment is
\begin{align}
    \frac{ d \langle f_{e} \rangle}{d\mll\, d\hll}
             = & \frac{\mll}{6s} \int_{-1}^{1}\,dc_\theta\,
             f_{e}(c_{\theta}) \left[
             G_{q\bar{q}}(\mll,\hll,Q^2)
             +
             G_{\bar{q}q}(\mll,\hll,Q^2)
             \right]
             \frac{d\hat{\sigma}}{d\hat{c}_\theta}(c_\theta,\mll),
             \label{eq:even-moment}
\end{align}
while for odd functions of $c_\theta$, i.e., $f_o(-c_{\theta}) = -f_o(c_{\theta})$, 
\begin{align}
    \frac{ d \langle f_o \rangle}{d\mll\, d\hll}=
            \xi\frac{\mll}{6s} \int_{-1}^{1}\,dc_\theta\,
             f_o(c_{\theta}) \left[
             G_{q\bar{q}}(\mll,\hll,Q^2)
             -
             G_{\bar{q}q}(\mll,\hll,Q^2)
             \right]
             \frac{d\hat{\sigma}}{d\hat{c}_\theta}(c_\theta,\mll),
             \label{eq:odd-moment}
\end{align}
where we have made the additional variable change $c_\theta\to\xi c_\theta$ and exploited the symmetry properties of the weight functions, $f_{e,o}$. The moments select the even and odd parts of the $c_\theta$ dependence of the differential cross section, respectively.

\subsection{Obtaining $B_0$ and $D_0$}\label{subsec:B0}
We are primarily interested in the $\tilde{B}_0$ and $\tilde{D}_0$ coefficients associated to the $Y^0_3(c_\theta)$ and $Y^0_4(c_\theta)$ spherical harmonics, as the other $\tilde{B}$ and $\tilde{D}$ ($Y^{\pm i}_{3,4}(c_\theta, \phi)$ with $i = 1,2,3, {\rm \, and~ subscript}~ 4 {\rm \; for \; } \tilde{D}$) first appear at $\mathcal{O}(\alpha_s)$~\cite{Alioli:2020kez}.
The coefficient $\tilde{B}_{0}$ in Eq.~\eqref{eq:ang_para} is extracted by
\begin{align}
\tilde{B}_{0}(\mll,\hll)= \sqrt{\frac{16\pi}{7}} \cdot \frac{14}{3}
\frac{
\int_{-1}^{1}dc_{\theta}
\frac{d \sigma_{pp \rightarrow \ell^- \ell^+}}{d \mll d \hll dc_{\theta}}\cdot Y_3^0(c_\theta)}
{\frac{d \sigma_{pp \rightarrow \ell^- \ell^+}}{d \mll d \hll}}.
\label{eq:op28}
\end{align}
with $Y_3^0(c_\theta) =\sqrt{\frac{7}{16\pi}}(5c^3_\theta-3 c_\theta)$. Here, the factor of $2\pi$ from the azimuthal integration is absorbed in the definition of the differential cross section. In general, both the numerator and denominator of Eq.~\eqref{eq:op28} depend on the Wilson coefficients of interest, and the denominator, especially, can potentially include effects from operators not contributing to the $l\geq 3$ spherical harmonics. We find it more convenient to define a new $B_{0}$ which is not normalized by a differential cross section: 
\begin{align}
\frac{d B_{0}}{d \mll d \hll} =\sqrt{\frac{16\pi}{7}}\cdot \frac{14}{3} \int_{-1}^{1}dc_{\theta} \ \frac{d \sigma_{pp \rightarrow \ell^- \ell^+}}{d \mll d \hll dc_{\theta}} \cdot Y_3^0(c_\theta).\label{eq:op29}
\end{align}
Hereafter, we refer to the new, un-normalised moments as $B_0$ and $D_0$. 
Using Eqs.~\eqref{eq:odd-moment} and~\eqref{eq:even-moment}, we obtain the master formulae to compute $B_0$ and $D_0$, respectively:
\begin{align}
    \frac{dB_0}{d\mll\, d\hll} &=  \frac{4\sqrt{7\pi}}{9}\cdot \frac{\xi\mll}{s} \int_{-1}^{1}\,dc_\theta\,
             Y_3^0(c_\theta) \left[
             G_{q\bar{q}}(\mll,\hll,Q^2)
             -
             G_{\bar{q}q}(\mll,\hll,Q^2)
             \right]
             \frac{d\hat{\sigma}}{d\hat{c}_\theta}(c_\theta ,\mll).
    \label{eq:B0eq3}\\
    \frac{d D_0}{d\mll\, d\hll} &=  \frac{\sqrt{\pi}}{3}\cdot\frac{\mll}{s} \int_{-1}^{1}\,dc_\theta\,
             Y_4^0(c_\theta) \left[
             G_{q\bar{q}}(\mll,\hll,Q^2)
             +
             G_{\bar{q}q}(\mll,\hll,Q^2)
             \right]
             \frac{d\hat{\sigma}}{d\hat{c}_\theta}(c_\theta ,\mll).
    \label{eq:D0eq}
\end{align}
Since the differential cross section is a polynomial function of $c_{\theta}$, $B_0$ and $D_0$ select terms with even and odd powers, respectively.

Finally, we remark that the Breit-Wheeler process $\gamma \gamma \to e^- e^+$ (and  $\gamma \gamma \to \mu^- \mu^+$ in our case) occurs at the LHC from the non-zero probability of finding a photon in the proton. This process has been studied as part of the EW corrections to the SM angular coefficients, $A_i$~\cite{Frederix:2020nyw}, and was found to have some impact on the predictions.
Its LO, partonic angular differential cross section in the centre of mass frame is given by:
\begin{align}
    \frac{d\hat{\sigma}}{d\hat{c}_\theta}(\hat{c}_\theta) &=
    \frac{2\pi \alpha^2}{\hat{s}}\beta
    \left( \frac{1 + \beta^2 \hat{c}_\theta^2}{1 - \beta^2\hat{c}_\theta^2}
    + 
     \frac{2m^2_\ell/E^2_\gamma}{1 - \beta^2\hat{c}_\theta^2}
    -
      \frac{2m^4_\ell/E^4_\gamma}{(1 - \beta^2\hat{c}_\theta^2)^2}
    \right),
\end{align}
where $E_\gamma$ is the energy of the initial photon, $\hat{c}_\theta$ is defined between an incoming photon and the outgoing negatively charged lepton, $\beta = |\vec{k}|/E_\gamma$, $\vec{k}$ is the momentum of the negatively charged lepton in the center-of-mass frame, and $m_\ell$ is its mass. Since the initial state is charge-symmetric, the angular dependence is symmetric in $\hat{c}_\theta$ and it therefore cannot contribute to $B_0$, while it does contribute to $D_0$. Our work is focused on obtaining sensitivity projections, such that we always assume that any irreducible SM component can be predicted and subtracted from the observed quantity with a rather small error compared to the dominant statistical and systematic uncertainties. 
In the invariant mass region of $100< m_{\ell\ell}< 1000~$~GeV that we are interested in, the LO total cross section from the Breit-Wheeler process is 11.8~pb, while the Drell-Yan process gives 151.8~pb. Applying typical experimental kinematic selections on the leptons of $p_T > 25$~GeV and $|\eta| < 2.4$, the cross sections become 0.8~pb and 68.9~pb respectively, reducing the Breit-Wheeler process down to a percent-level effect. These numbers were obtained with \verb|MadGraph5_aMC@NLO|~\cite{Alwall:2014hca}, using the  \verb|NNPDF31_nlo_as_0118_luxqed| PDF set (lhaid=324900)~\cite{Bertone:2017bme},
which features an inelastic photon PDF matched to an elastic photon PDF at low momentum transfer using the LUXqed formalism~\cite{Pascoli:2018heg,Manohar:2016nzj,Manohar:2017eqh}.

\section{High-luminosity LHC sensitivity to the cutoff scale}\label{sec:cutoff}
\begin{figure}[h]
	\centering
		\includegraphics[width=.45\linewidth]{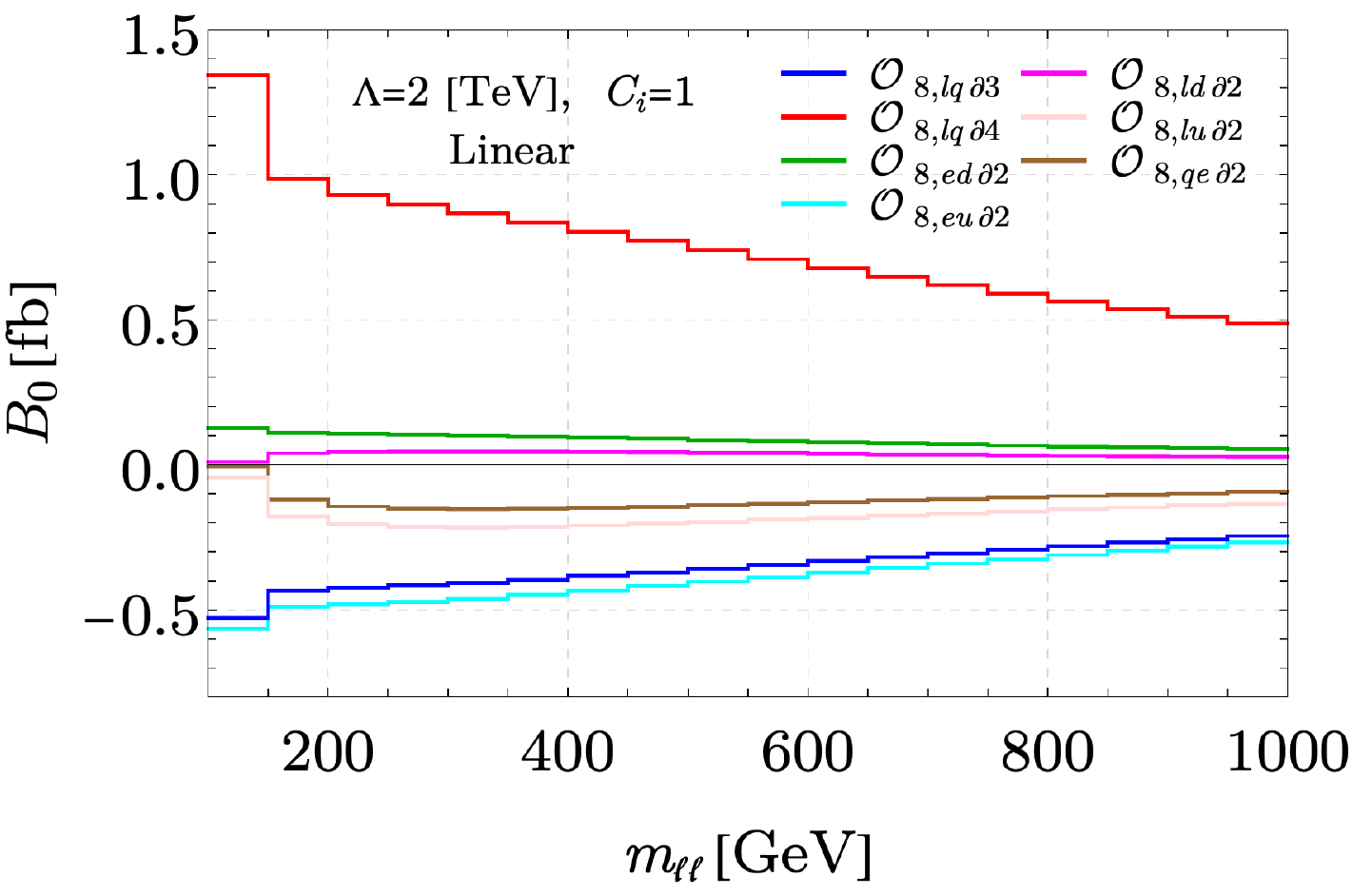}
        \includegraphics[width=.45\linewidth]{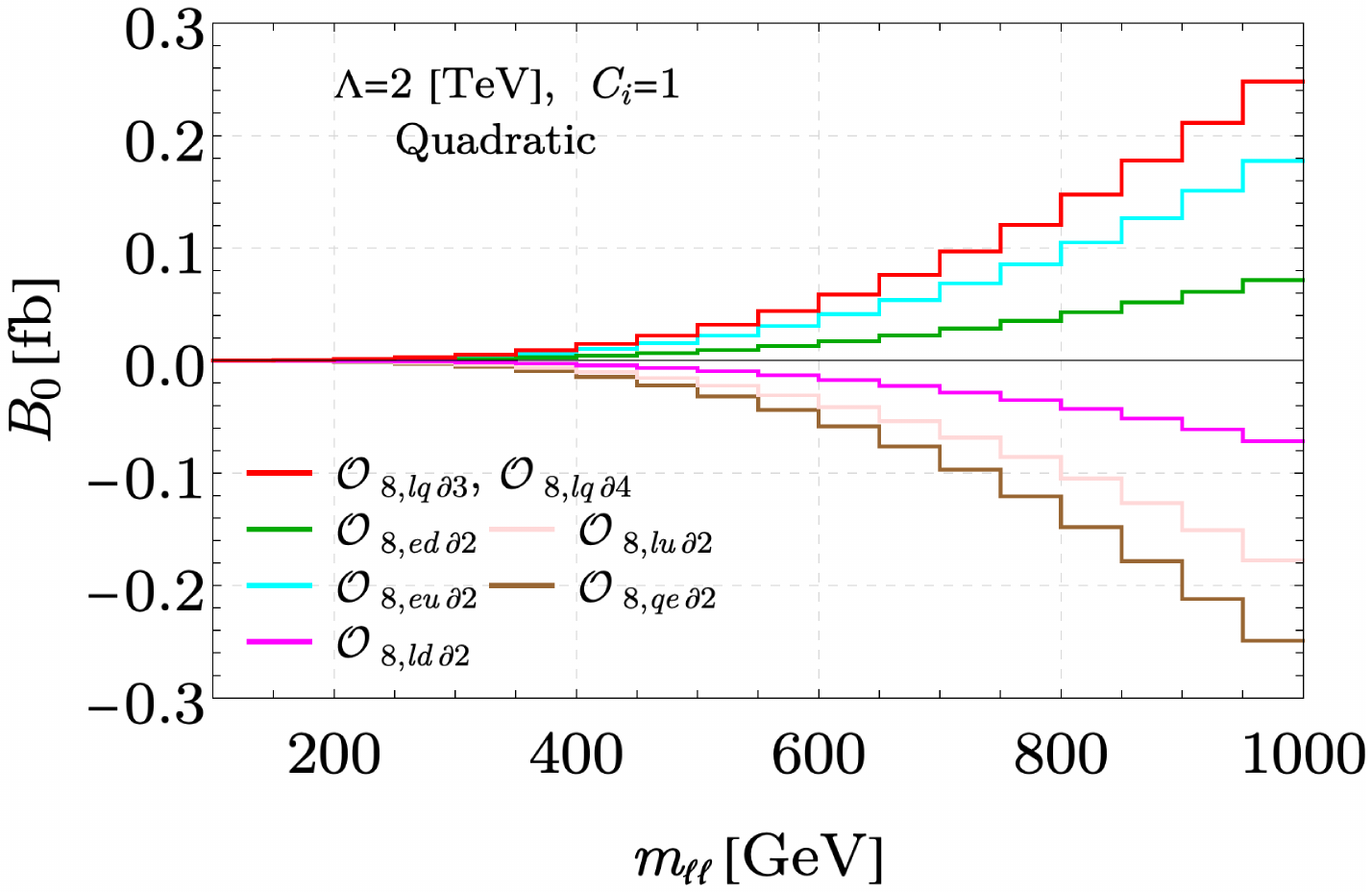}
        \includegraphics[width=.45\linewidth]{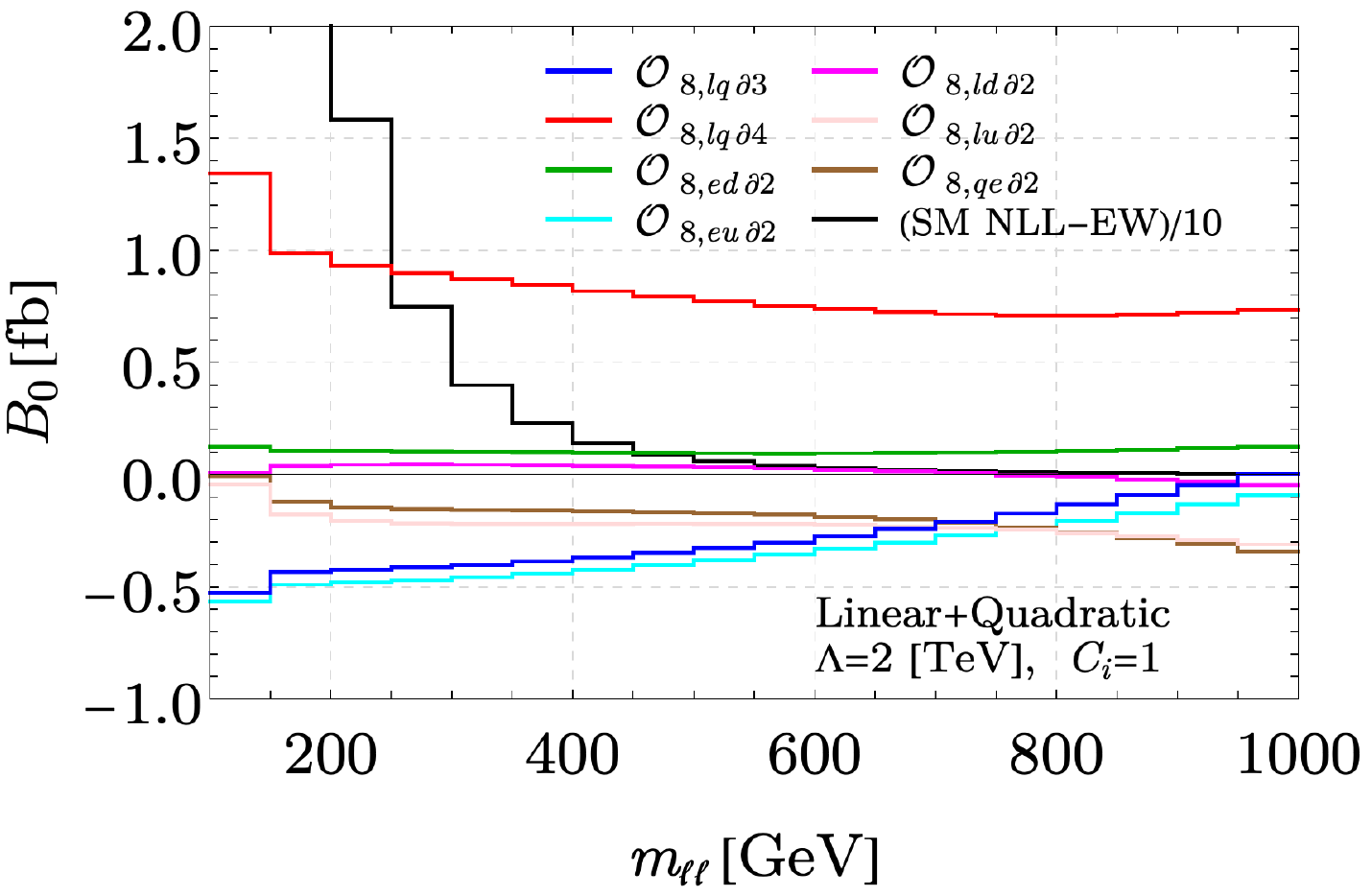}
        \includegraphics[width=.45\linewidth]{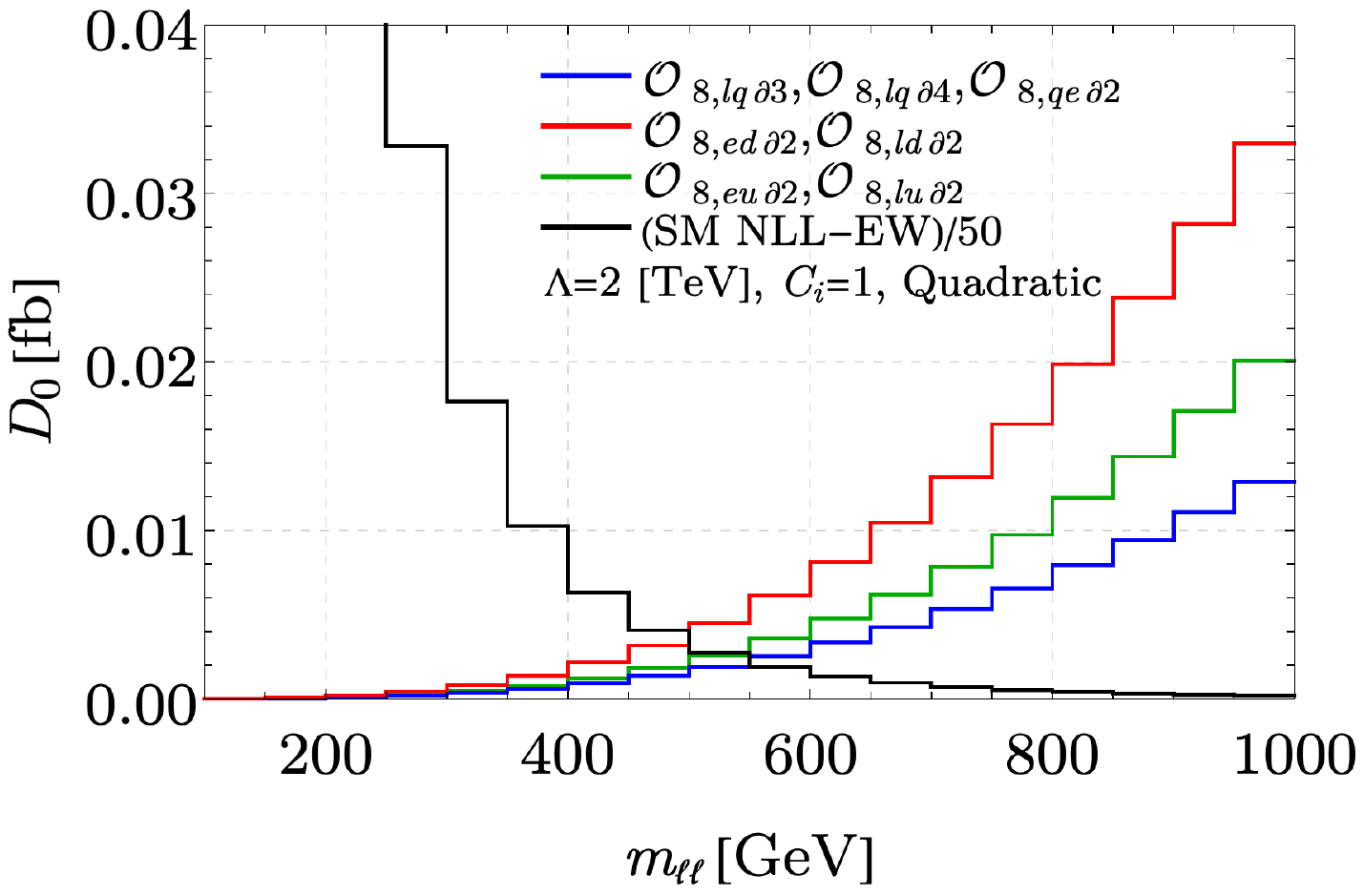}
	\caption{Individual operator contributions to the differential distributions for the $B_{0}$ and $D_0$ angular moments, setting $\Lambda=2$ TeV and the Wilson coefficient $C_i=1$. \emph{Upper left}: $B_0$ contributions linear in $C_i$.  \emph{Upper right}: $B_0$ contributions quadratic in $C_i$. \emph{Lower left}: Sum of the linear and quadratic contributions in $C_i$ to $B_0$. The SM NLL EW prediction is also shown, scaled by a factor 0.1 \emph{Lower right}: $D_0$ contributions quadratic in $C_i$ (there are no linear contributions). The SM NLL EW prediction is also shown, scaled by a factor 0.02.}
  \label{fig:B01000}
\end{figure}

We now turn to determining the sensitivity of our differential angular observables to the Wilson coefficients at the high Luminosity LHC (HL-LHC). 
We therefore take a collider centre of mass energy, $\sqrt{s}=14$ TeV, and fix the cutoff to $\Lambda=2$ TeV. We use the \verb|NNPDF31_nlo_as_0118_luxqed| PDFs. We use Eq.~\eqref{eq:op29} to obtain differential distributions for the angular moments.
The top row of Figure~\ref{fig:B01000} shows the linear and quadratic contributions of each operator to the differential $B_{0}$ distribution in $\mll$ bins of 50 GeV from 100 to 1000 GeV, for a Wilson coefficient value of 1. 
The lower left panel shows the contributions from the full dim-8 amplitude (linear plus quadratic) of a given operator.  The lower right panel shows the contribution of each operator to the  $l=4$ moment, $D_0$, which only arises at quadratic level. The aforementioned SM NLL EW contribution is also shown for reference. Apart from the mild shape differences at low invariant mass from interference with the $Z$-pole, we can see the common, characteristic energy growth with respect to the SM, which is even more pronounced at the quadratic level. Since all operators have the same high-energy dependence, \emph{a-priori} they cannot be distinguished in the $\mll$ distribution of $B_0$ at the linear level. However, the relative magnitude and sign of their linear and quadratic contributions are not always the same and the (positive-definite) $D_0$ moment also brings further discrimination power.

Table~\ref{tab:dependence} illustrates the salient features of the angular dependence of each operator contribution to the partonic channels of Drell-Yan and its angular moments. 
It summarises the information in Eqs.~\eqref{eq:deltaM_lq3}-\eqref{eq:dim8_M_squared_end}, in the $\sqrt{\hat{s}}\gg M_Z$ limit. The numerical coefficients in front of the interference contributions are obtained by inserting the SM quantum numbers into the corresponding expression, neglecting the common factors of $8\pi\alpha \hat{s}^2$. They give an indication of the relative importance of the interference contribution of each operator at high invariant mass.
\begin{table}[t!]
\centering
\renewcommand\arraystretch{1.2}
	\setlength{\tabcolsep}{1.mm}
\begin{tabular}{c c r c c c c}
\hline \hline Operators & Quark & \multicolumn{1}{c}{$\Delta |\mathcal{M}^{(8)}|^2 $} & $|\mathcal{M}^{(8)}|^2 $ & $B_0 {\rm ( lin.)}$ & $B_0 {\rm (quad.)}$& $D_0 {\rm (quad.)}$\\
\hline \hline $\mathcal{O}_{8, lq\partial 3}+\mathcal{O}_{8, lq\partial 4}$ & $d\bar{d}$ & $ 1.9 \cdot c_{\theta}(1+c_{\theta})^2$  & $c_{\theta}^2(1+c_{\theta})^2$ & $+$ & $+$& $+$\\
 $\mathcal{O}_{8, lq\partial 3}-\mathcal{O}_{8, lq\partial 4}$ &$u\bar{u}$ & $-2.4 \cdot c_{\theta}(1+c_{\theta})^2$  & $c_{\theta}^2(1+c_{\theta})^2$ & $-$ & $+$& $+$\\
 $\mathcal{O}_{8, ed\partial 2}$ & $d\bar{d}$ & $0.43 \cdot c_{\theta}(1+c_{\theta})^2$  & $c_{\theta}^2(1+c_{\theta})^2$   & $+$ & $+$ &$+$\\
 $\mathcal{O}_{8, eu\partial 2}$ & $u\bar{u}$ & $-0.87  \cdot c_{\theta}(1+c_{\theta})^2$  & $c_{\theta}^2(1+c_{\theta})^2$ &  $-$ & $+$&$+$\\
 $\mathcal{O}_{8, ld\partial 2}$ & $d\bar{d}$ & $0.22 \cdot c_{\theta}(1-c_{\theta})^2$  & $c_{\theta}^2(1-c_{\theta})^2$   & $+$& $-$&$+$\\
 $\mathcal{O}_{8, lu\partial 2}$ & $u\bar{u}$ & $-0.43 \cdot c_{\theta}(1-c_{\theta})^2$  & $c_{\theta}^2(1-c_{\theta})^2$  & $-$ & $-$&$+$\\
 $\mathcal{O}_{8, qe\partial 2}$ & $u\bar{u}$ or $d\bar{d}$ & $-0.22 \cdot c_{\theta}(1-c_{\theta})^2$  & $c_{\theta}^2(1-c_{\theta})^2$  & $-$ & $-$ &$+$\\
\hline \hline
\end{tabular}
\caption{
\label{tab:dependence}
Schematic angular dependence of the linear and quadratic contributions of the dim-8 operators to $q\bar{q}\to\ell^-\ell^+$ scattering in the $\sqrt{\hat{s}}\gg M_Z$ limit. We also show the relative sign of the induced $B_0$ and $D_0$ after integrating with the spherical harmonic functions $Y_3^0$ and $Y^0_4$. The ``Quark'' column denotes quark-anti quark channel(s) mediated by each operator.}
\end{table}
The table highlights operators that we expect to be distinguishable, due to either having different combinations of signs in their linear and quadratic contributions to $B_0$ and $D_0$, or by mediating different combinations of the $u\bar{u}$ and $d\bar{d}$ initial states. Although not evident in the invariant mass distributions, specific combinations of partonic initial states leave an imprint in the $\hll$ distribution, via the respective PDFs.  We can isolate combinations of operators that mediate specific initial states in all cases apart from $C_{8,qe\partial 2}$, which contributes to both with -- coincidentally -- the same numerical prefactor in the interference term. Furthermore, the relative size of the linear and quadratic terms will modify the shape of the $\mll$ spectra for a given value of the Wilson coefficient, which may provide some distinguishing power at intermediate invariant masses. Finally, the mild differences in the $\mll$ spectra near the $Z$-pole may confer a modest amount of additional information. 

We see that it may be challenging to disentangle $C_{8,lq\partial3}+C_{8,lq\partial4}$ from $C_{8,ed\partial2}$ and $C_{8,lq\partial3}-C_{8,lq\partial4}$ from $C_{8,eu\partial2}$, using only $B_0$ and $D_0$, since they yield the same combination of signs and mediate the same initial state. However, they do predict significantly different relative importance of the interference term. A similar argument can be applied to $C_{8,lu\partial2}$ and $C_{8,qe\partial2}$; even though they do not mediate the same initial state in this case, they appear difficult to distinguish in Figure~\ref{fig:B01000} for $\mathcal{O}(1)$ coefficients. Conversely, the operators that have totally opposite signs in their $B_0$ contributions may lead to cancellations in specific directions of the parameter space. This may lead to relatively weakly constrained directions and increase the importance of the $D_0$ observable, which is positive definite. We point out some instances of this interplay later on in this Section and in Section~\ref{sec:positivity_v}. 

 In our analysis, motivated by the desire to maximally distinguish the different operator contributions using the angular moments, their energy dependence, and the sensitivity of $\hll$ to partonic initial states, we take a 10$\times$10 rectangular binning of our $(\mll,\hll)$ double-differential distributions with the following bin edges:
\begin{align*}
    \mll&:\,
    { \{100,~190,~280,~370,~460,~550,~640,~730,~820,~910,~1000 \}} 
    \,\text{GeV},\\
    \hll&:\,
    { \{ -5,~-4,~-3,~-2,~-1,~0,~1,~2,~3,~4,~5\},}
\end{align*}
and obtain the dimension-8 operator contributions as a function of the Wilson coefficients, $\vec{C}$, up to quadratic order. We limit the maximum energy scale to 1 TeV in an attempt to mitigate the potential sensitivity to unknown effects from higher-dimensional operators, which we further discuss at the end of this section. Each $(\mll,\hll)$ bin will have pair of observables $\vec{O}^i=(B^i_0,D^i_0)$ associated to it, with which we construct a binned $\chi^2$ formula. The likelihood depends on the predicted and observed values of $\vec{O}^i$, denoted by $\vec{O}^i_{\vec{C}}=\left(B^i_0(\vec{C}),D^i_0(\vec{C})\right)$ and $\vec{O}^i_{\vec{C}_0}=\left(B^i_0(\vec{C}_0),D^i_0(\vec{C}_0)\right)$, respectively, and takes the standard form:
\begin{align}
\chi^2(\vec{C},\vec{C}_0)=\sum_{i}
(\vec{O}^i_{\vec{C}}-\vec{O}^i_{\vec{C}_0})^\top\cdot \mathbf{V}^{-1}\cdot (\vec{O}^i_{\vec{C}}-\vec{O}^i_{\vec{C}_0}),
\end{align}
where $i$ sums over each $(\mll,\hll)$ bin. $C_0$ denotes the `true' parameter space point that is observed, which we assume to be the SM, $\vec{C}_0=\vec{0}$, unless otherwise stated. In this case, $\vec{O}^i_{\vec{C}_0}=0$, meaning that the $\chi^2$ and $\Delta\chi^2=\chi^2-\chi^2_\mathrm{min.}$ formulae coincide, since the minimum of the $\chi^2$ is at $\chi^2_\mathrm{min.}=0$. Constraints on the parameter space are then extracted via:
\begin{align}
\Delta\chi^2(\vec{C})=\chi^2(\vec{C},\vec{0}) \leq 3.84.
\end{align}
The critical value of 3.84 corresponds to the 95\% C.L. allowed region for a $\vec{C}$ comprised of one degree of freedom, and should be replaced by the appropriate value for the dimensionality of the parameter space at hand.  
In a given bin, $B_{0i}$ and $D_{0i}$ are statistically correlated since they both depend on the underlying angular distribution. $\mathbf{V}$ denotes the associated covariance matrix\footnote{See Appendix~\ref{app:cov} for a derivation of the covariance matrix for a set of such moment functions.}, whose entries are calculated for each bin as follows:
\begin{align}
\nonumber
V_{ij}&= \frac{1}{L} 
\int_{m_\text{min.}}^{m_\text{max.}} d\mll
\int_{\eta_\text{min.}}^{\eta_\text{max.}} d\hll
\int_{-1}^{1} dc_{\theta}
\frac{d \sigma_{pp\to \ell^- \ell^+}}{d\hll ~d\mll~dc_{\theta}} 
\cdot F_{ij}(c_\theta),\\
F_{11}&=\frac{448\pi}{9}\left(Y_3^0(c_{\theta})\right)^2;\quad
F_{22}=\frac{36 \pi^3}{49}\left(Y_4^0(c_{\theta})\right)^2;\quad
F_{12}=F_{21}=\sqrt{\frac{16}{7}} 4 \pi^2 Y_3^0(c_{\theta})Y_4^0(c_{\theta}),\label{eq:covij}
\end{align}
with $(m_\text{min.},m_\text{max.})$ and $(\eta_\text{min.},\eta_\text{max.})$ being the bin boundaries in $\mll$ and $\hll$, respectively, and $L$ being the integrated luminosity of the collider, which we set to be $3000$ fb$^{-1}$, the High Luminosity LHC (HL-LHC) target. When the weight function is manifestly even in $c_\theta$, such as for the individual variances (diagonal entries of $\mathbf{V}$) the contribution is dominated by the total cross section and $A_0$ moment contributions to the angular distributions from the SM. We compute these contributions at NLO QCD accuracy with \verb|MadGraph5_aMC@NLO|, and find the EW corrections to be negligible. We note that if one assumes the observation of non-zero Wilson coefficients, the covariance matrix will receive sub-dominant contributions from the SMEFT, not only from the dim-8 operators at hand but also potentially from \emph{a priori} unknown dim-6 operators that we have not considered in this work. We do not expect these to qualitatively change our results, but stress that their impact should be assessed in a more comprehensive analysis incorporating not only other angular moments but also low-energy data in the spirit of Ref.~\cite{Boughezal:2021kla}. We leave this interesting possibility to future work.

\begin{figure}[h!]
	\centering
		\includegraphics[width=.7\linewidth]{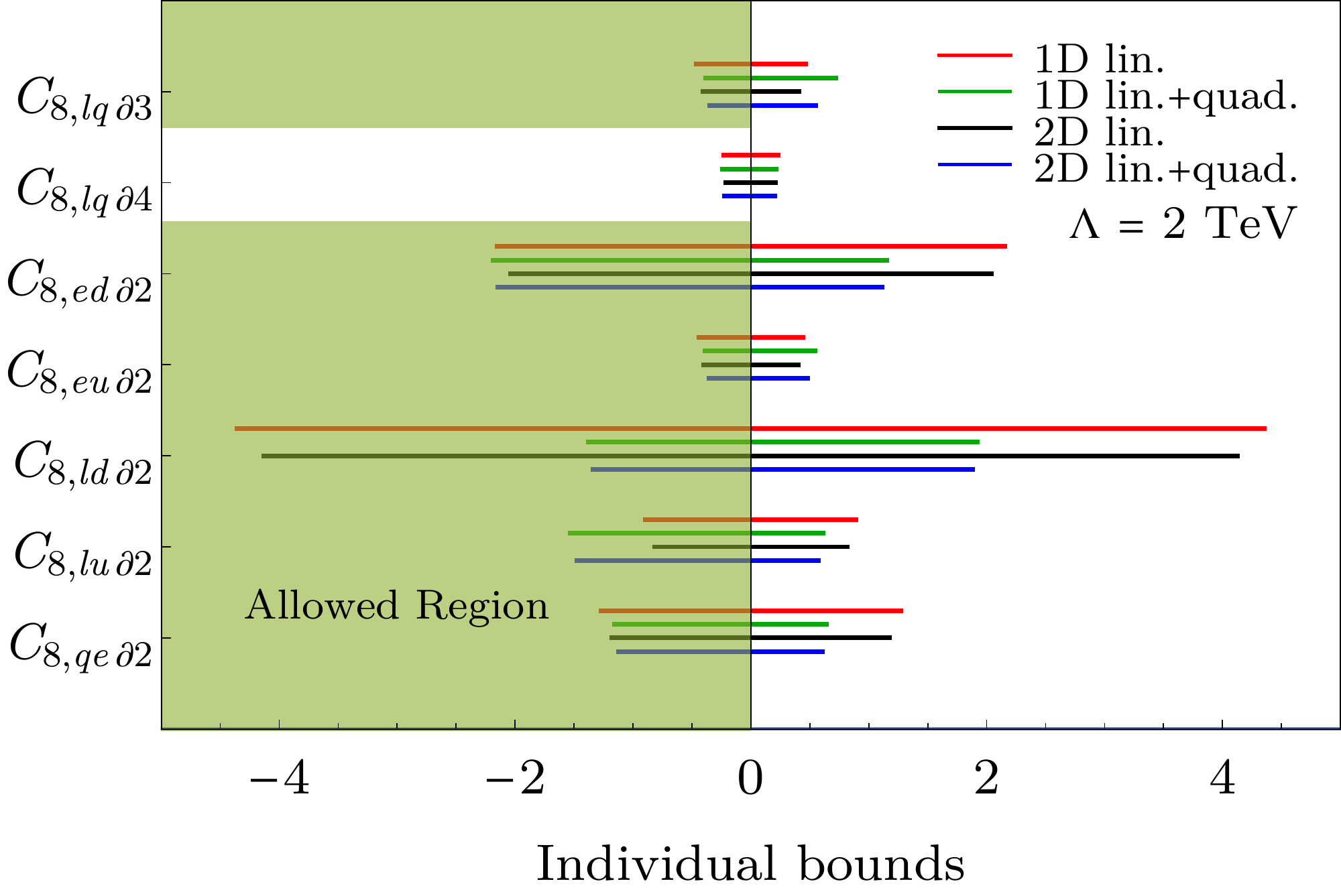}
	\caption{Individual bounds on the dimension-8 coefficients for $\Lambda=2$ TeV. 95\% confidence intervals are given at both the linear and quadratic levels and obtained with either the 1D or 2D differential $B_0$ analyses, see text for details. ``Allowed Region'' indicates the elastic positivity bounds of Table \ref{tab:elastic_bounds}. }
	\label{fig:constrainsSM}
\end{figure}

We perform the $\chi^2$ analysis in both the 1D and 2D cases, taking either the single- or double-differential $m_{\ell \ell}$ and (${m_{\ell \ell},\eta_{\ell \ell}}$) distributions for $B_0$, respectively. In the 1D case, we use 10 bins in $\mll$ only, integrating over $\hll$. Moreover, we compare bounds obtained when including the dimension-8 operator contribution to the differential cross section at the linear and quadratic levels, \emph{i.e.}, by including only the second, or both the second and third terms of Eq.~\eqref{eq:ampsq}. The individual (setting all other coefficients to zero), 95\% Confidence Level (C.L.) bounds on various Wilson coefficients are shown in Figure~\ref{fig:constrainsSM} and their numerical values are given in Table~\ref{tab:individualbounds}.
\begin{table}[ht]
\centering
\renewcommand\arraystretch{1.2}
	\setlength{\tabcolsep}{2.6mm}
\begin{tabular}{cccccc}
\hline \hline Coeff. & 1D lin. & 1D lin.+quad. & 2D lin. & 2D lin.+quad. &+$D_0$ \\
\hline \hline
 $C_{8, lq\partial 3}$ & $(-0.46, 0.46)$ & $(-0.38, 0.72)$ & $(-0.41, 0.41)$ & $(-0.35, 0.55)$ & $(-0.32, 0.44)$\\
 $C_{8, lq\partial 4}$ & $(-0.23, 0.23)$ & $(-0.24, 0.22)$ & $(-0.21, 0.21)$ & $(-0.22, 0.20)$ & $(-0.20, 0.18)$\\
 $C_{8, ed\partial 2}$ & $(-2.2 , 2.2 )$ & $(-2.2 ,	1.2 )$ & $(-2.0	, 2.0 )$ & $(-2.1 , 1.1 )$ & $(-1.8 , 1.1 )$\\
 $C_{8, eu\partial 2}$ & $(-0.44, 0.44)$ & $(-0.39,	0.55)$ & $(-0.40, 0.40)$ & $(-0.36, 0.48)$ & $(-0.33, 0.41)$\\
 $C_{8, ld\partial 2}$ & $(-4.4 , 4.4 )$ & $(-1.4 , 1.9 )$ & $(-4.1 , 4.1 )$ & $(-1.3 , 1.9 )$ & $(-1.1 , 1.5 )$\\
 $C_{8, lu\partial 2}$ & $(-0.89, 0.89)$ & $(-1.5 , 0.61)$ & $(-0.81, 0.81)$ & $(-1.5 , 0.57)$ & $(-1.2 , 0.5 )$\\
 $C_{8, qe\partial 2}$ & $(-1.3 , 1.3 )$ & $(-1.2 ,	0.64)$ & $(-1.2 , 1.2 )$ & $(-1.1 , 0.61)$ & $(-0.93, 0.54)$\\
\hline \hline
\end{tabular}
\caption{Individual bounds on the dimension-8 coefficients for $\Lambda = 2$~TeV at the linear and quadratic levels obtained with the 1D and 2D differential $B_0$ analyses, as shown in Fig.~\ref{fig:constrainsSM}. The last column shows the impact of including $D_0$ in the 2D linear+quadratic case.}
\label{tab:individualbounds}
\end{table}
We see that, at the individual level, bounds of order 1 or better can be obtained for most of the operators. The exceptions are the two operators that only mediate the $d\bar{d}$ initial state and therefore have a suppressed overall contribution that leads to bounds of order 2--4. Furthermore, the addition of the $\hll$ information only mildly improves the sensitivity, which is expected since it is mainly included to offer distinguishing power that is not useful when looking at operators one at a time. The impact of quadratic terms is more evident, especially in the more weakly constrained directions, since larger coefficient values are allowed. Their inclusion generically leads to a tightening of the bounds, at most by a factor 2.7. Table~\ref{tab:individualbounds} also shows the mild impact of including $D_0$, which improves the sensitivity by 10-20\%. Assuming, instead, Wilson coefficient values of one, the scales probed range between 1.7 and 3 TeV. 

The distinguishing power of our observables can be quantified by considering profiled results, allowing all operators to float simultaneously in the fit. Indeed, we find that without the inclusion of the 2D information, the quadratic dim-8 effects, or the $D_0$ angular moment, flat directions exist that prevent the extraction of reliable confidence intervals. Including the full information, however, does allow us to obtain profiled bounds, which are given in Table~\ref{tab:marginalised}. They correspond to a reduced sensitivity to a scale of 1.7-2 TeV when assuming $C_i=1$, as expected for a more global analysis.
\begin{table}[ht]
\centering
\renewcommand\arraystretch{1.2}
	\setlength{\tabcolsep}{0.7mm}
\begin{tabular}{|c| c| c| c |c |c |c| c|}
\hline  Coeff. & $C_{8, lq\partial 3}$ & $C_{8, lq\partial 4}$ & $C_{8, ed\partial 2}$ & $C_{8, eu\partial 2}$ & $C_{8, ld\partial 2}$ &  $C_{8, lu\partial 2}$ &$C_{8, qe\partial 2}$\\
\hline $B_0+D_0$ & $(-1.1,	1.1)$ & $(-0.95,	0.85)$ & $(-1.8,	1.8)$  & $(-1.2,	1.3)$ & $(-1.7,	1.7) $ & $(-1.2,	1.2)$  & $(-1.0,	1.0)$  \\
\hline 
\end{tabular}
\caption{ Profiled bounds on the dimension-8 coefficients for $\Lambda = 2$~TeV at the quadratic level obtained with the 2D differential information in $B_0$ and $D_0$.}
\label{tab:marginalised}
\end{table}
This shows that the system is somewhat under-constrained, requiring, in particular, the inclusion of $1/
\Lambda^8$ effects to effectively distinguish different operator contributions. When we neglect these terms, the Fisher information matrix has 2 relatively well constrained eigenvectors, and a third moderately well constrained one. Their coefficients are bounded at values of 0.15, 2.5 and 40, respectively (again, assuming $\Lambda=2$ TeV), while the other four directions are essentially unconstrained. This means that care should be taken in interpreting the results, since we have not considered all possible effects arising at dimension greater than 8.

\begin{figure}[ht!]
	\centering
		\includegraphics[width=\textwidth, trim=20 15 55 55, clip]{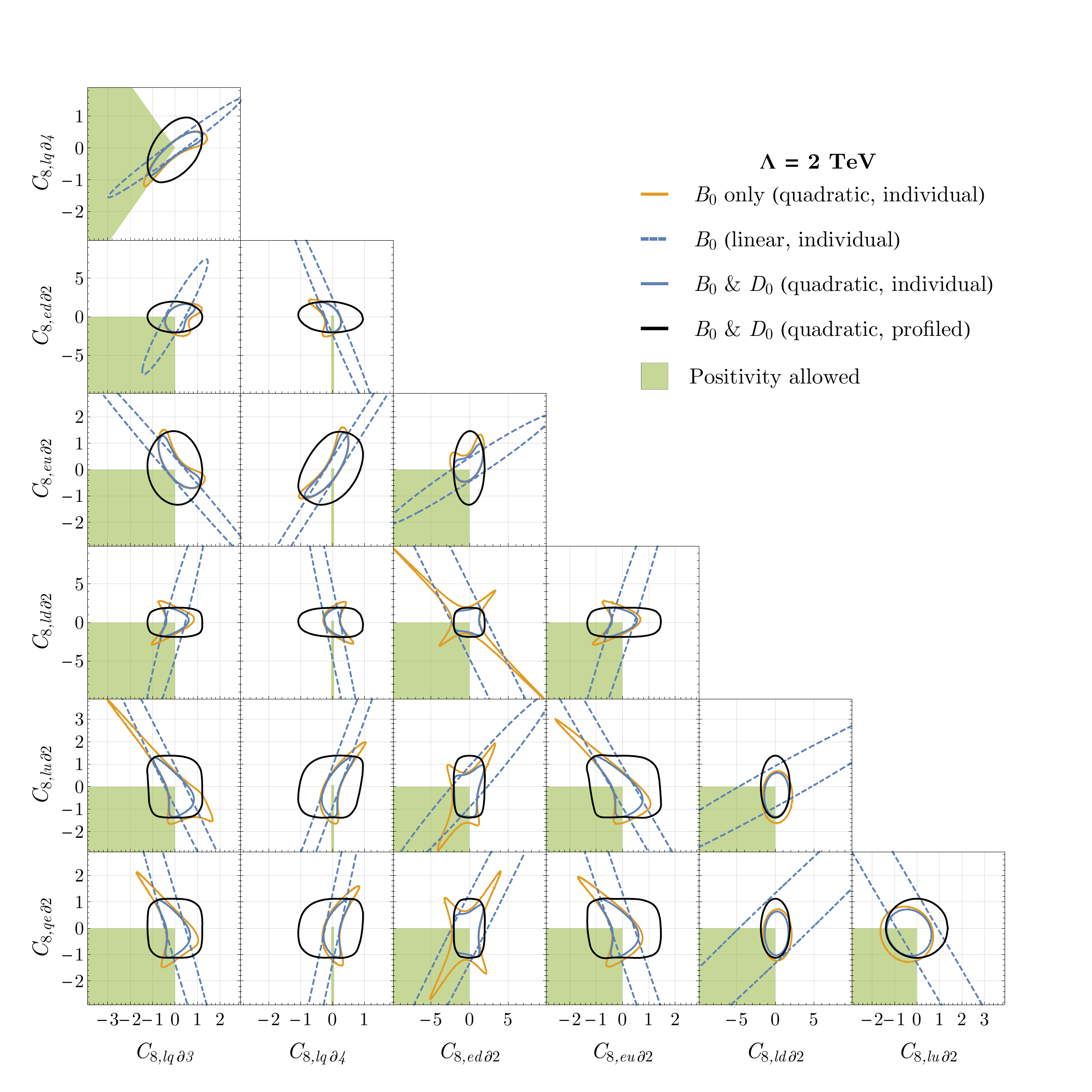}
	\caption{Two-dimensional individual, 95\% confidence regions in the dimension-8 coefficient space using the double-differential distributions of the $B_0$ and $D_0$ angular moments. The elastic positivity bounds of Table \ref{tab:elastic_bounds} are shaded in green.}
	\label{fig:constrainsSM2D}
\end{figure}
In order to explore the correlations between the different operators, we also study the constraints in 2D sub-spaces of the Wilson coefficients, again assuming the SM hypothesis is observed. Figure~\ref{fig:constrainsSM2D} summarises the confidence regions obtained in different limits from our analysis, in the 21 combinations of operators, corresponding to isocontours of $\chi^2=5.99$.
Individual regions (setting all other coefficients to zero) are given when including only the $B_0$ observable (orange), both $B_0$ and $D_0$ (solid blue) and for $B_0$ neglecting the quadratic contributions (dashed blue). The corresponding profiled regions, using the full $B_0$ and $D_0$ information at quadratic level, are shown in black. For reference, we also shade in green the region allowed by the positivity bounds. The constraints in the linear-only limit clearly highlight the presence of flat directions in the corresponding likelihood, most of which are lifted when including quadratic effects in $B_0$. However, approximate flat directions still remain in, \emph{e.g.}, the  2D planes involving $C_{8,ed\partial 2}$. This operator is one of the combinations that mediates only the $d\bar{d}$ intial state, which explains why it has on of the poorest sensitivities overall. The three operators with which the blind directions are apparent, $C_{8,qe\partial 2}$, $C_{8,lu\partial 2}$ and $C_{8,ld\partial 2}$, are exactly those which predict an opposite sign in $B_0$ (See Table~\ref{tab:dependence}) and cancel in a particular pair of directions (hence the ``x'' shape, since the quadratic term is insensitive to the coefficient sign). The poorest constrained directions then approximately follow those of the linear-only contours, where the sign of the coefficients is such that these terms also cancel.    Ultimately, including $D_0$ closes these directions off, allowing for the final, marginalised regions to be reliably obtained. The possible degeneracies highlighted in the discussion of Table~\ref{tab:dependence} are therefore never exact, and including the full information allows us to resolve the parameter space. This suggests that the low/intermediate invariant mass bins do play a role in discriminating the effects of the seven operators.

Overall, we find that the LHC can bound $\mathcal{O}(1)$ coefficients for a new physics scale of $\Lambda=2$ TeV, using the full information from $B_0$ and $D_0$ including the $\mathcal{O}(1/\Lambda^8)$ effects from the operators of interest. As previously mentioned, this result relies on the inclusion of partial higher-order contributions in the SMEFT expansion, neglecting other possible contributions at $\mathcal{O}(1/\Lambda^{n\geq 6})$. These can come from the interference between dim-6 and dim-8 amplitudes, or between dim-10/12 amplitudes and the SM. We have neglected the former by assuming the relevant dim-6 operators can be adequately constrained elsewhere (See, \emph{e.g.}, Refs.~\cite{Falkowski:2017pss,Boughezal:2021kla}). The contribution of operators of dim-($n>8$) has not been computed, and hence our results should be interpreted with caution. In particular, the 4- and 6-derivative counterparts of our dim-8 operators also contribute to the $l\geq3$ angular moments and may alter the picture of constraints. That said, we have limited the maximum energy bins of our analysis to 1 TeV, in an attempt to partly mitigate these effects. Furthermore, these operators will also contribute to even higher moments than our dim-8 ones, which could be used to further reduce their impact. A full angular analysis up to higher energies would be required to obtain an accurate picture of the sensitivity. Our study is intended to be a proof of principle focused on the ability to probe the positivity cone and infer information about UV states. Ultimately, these results should form part of a more complete, global analysis including all relevant operators that is beyond the scope of this work.

\section{Testing positivity}\label{sec:positivity_v}
The principles used to derive the positivity bounds are well-established, at least for the energy scales we have probed so far, which makes positivity bounds very reliable, and there is no concrete reason to expect any violation at high energies. Nevertheless, nature, \emph{i.e.} the experimental data must have the ultimate say on whether they are satisfied. Here we shall entertain the possibility that the positivity bounds may be violated at scales not very far from those probed the colliders. That is, we shall test the axiomatic principles of the S-matrix by postulating their violations and the degree to which these can be confirmed experimentally. 

Following \cite{Fuks:2020ujk}, we shall postulate that a particular set of non-zero values of the Wilson coefficients, $\vec{C}_{0}$, are observed, and quantify the violation of positivity by computing
\begin{align}
-\Delta^{-4} &\equiv {\rm min}\left[\underset{\rm{processes}}{\rm{min}} \frac{1}{2}\frac{d^2M (0)}{ds^2}, 0\right] = \frac{\delta(\vec{C}_0)}{\Lambda^4},
\label{eq:Delta}
\end{align}
with
\begin{align}
\label{ddC8}
\delta(\vec{C}_0) \equiv {\rm min}&\left[  
 -4C_{8, lq\partial 3}+4C_{8, lq\partial 4} , -4C_{8, lq\partial 3}-4C_{8, lq\partial 4},
-4C_{8, ed\partial 2},-4C_{8, eu\partial 2}, \right. \nonumber\\ 
& ~\left. -4C_{8, ld\partial 2},-4C_{8, lu\partial 2},-4C_{8, qe\partial 2}
,0\right].
\end{align}
where the subtracted amplitude $M(0)$ is defined in Eq.~\eqref{defMijkl0} and ``processes" in the minimization denote the different elastic scattering processes in Table~\ref{tab:elastic_bounds}. This is of course just a crude estimate, which serves our purposes at this stage. Geometrically, in the space of the Wilson coefficients, these bounds represent the facets of a polyhedral convex cone. Eq.~\eqref{eq:Delta} can be visualized as evaluating a (scaled) distance between the prospective experimental data point, $\vec{C}_{0}$, and the closest facet, $\vec{C}_{0}$ being outside the cone if the positivity is violated. We will have $\Delta^{-1}=0$ or $\Delta=\infty$ if positivity is satisfied, and if it is violated we have $\Delta={\Lambda}/\sqrt[4]{|\delta{C_{\text{min}}}|}$, where $\delta{C_{\text{min}}}$ is the non-zero minimum found with Eq.~(\ref{ddC8}).
In short, $\Delta$ has dimensions of energy and will be taken as a proxy for the scale of positivity violation. 
\begin{figure}[ht!]
	\centering
	    \includegraphics[width=0.5\textwidth]{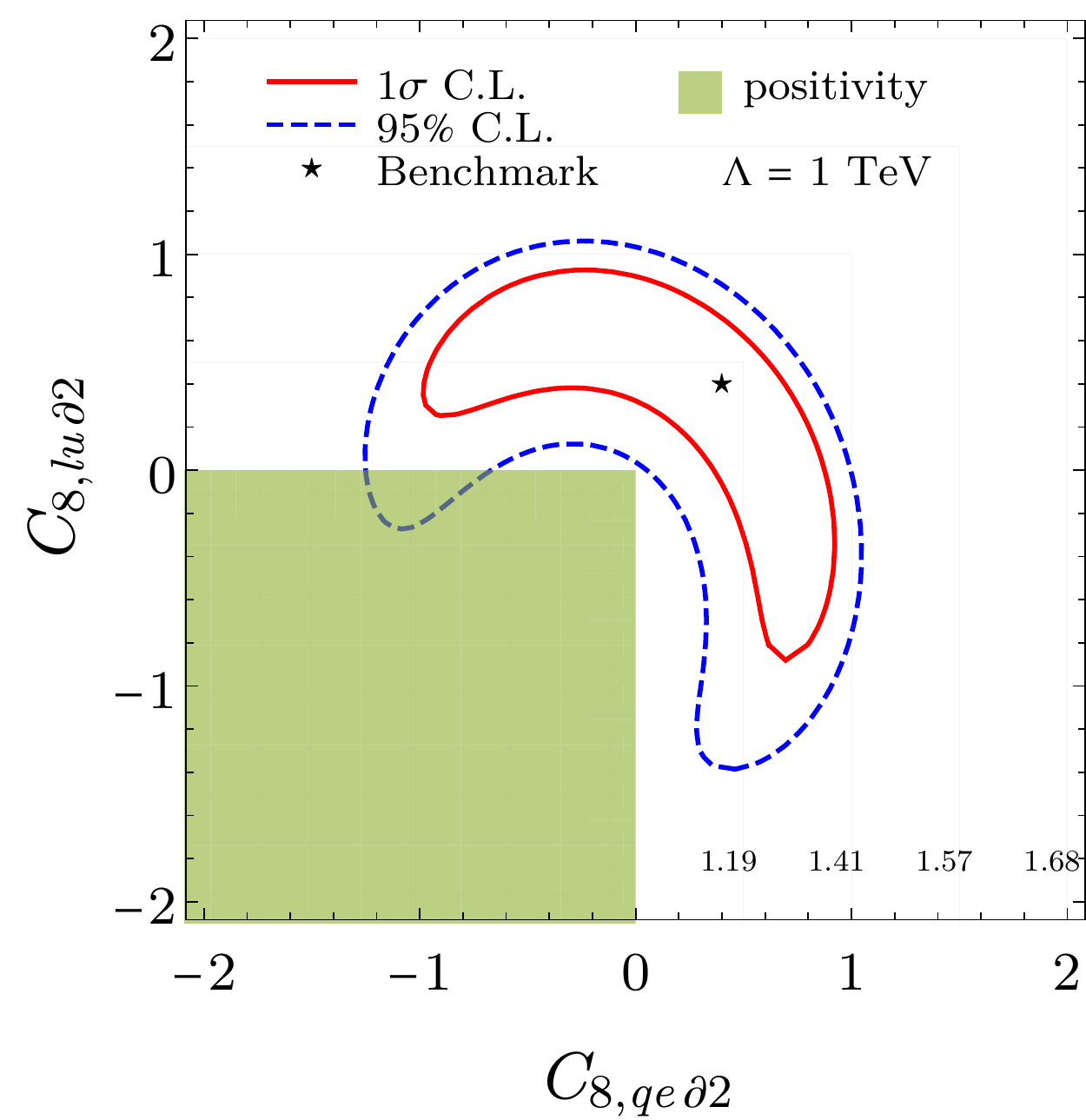}
	\caption{A benchmark case of positivity violation in the 2D subspace $(C_{8,qe\partial 2},C_{8,lu\partial 2})$ in the presence of experimental uncertainties. The green region is where the elastic positivity bounds are satisfied. The grey solid lines are the values of $\Delta^{-1}$. We set $\Lambda= 1$~TeV in this analysis.}
	\label{fig:violation2}
\end{figure}

Of course, experimentally, the Wilson coefficients can only be determined within some region at a given C.L., rather than pin-pointed to an exact location, so $\Delta$ can only be constrained in a range. Furthermore, the associated likelihood is a more complex function, since the covariance matrix now depends on $C_0$.  It is instructive to first see this in a simple example. To this end, we look at the 2D subspace $(C_{8,qe\partial 2},C_{8,lu\partial 2})$, where a benchmark $(0.4,0.4)$ is chosen for these two coefficients, and the other coefficients are set to zero. We set $\Lambda= 1$~TeV in this analysis. We confront this parameter space with the likelihood built from the double-differential HL-LHC data for $B_0$ and $D_0$ that we have obtained in the previous sections\footnote{We note that the presence of non-zero dim-8 coefficients would suggest (although not guarantee) the presence of dim-6 operators as well. These operators can contribute indirectly to the dim-8 likelihood through the covariance matrix entries of Eq.~\eqref{eq:covij}, although we stress that they are typically dominated by the SM cross section contributions in each bin.}. In Figure~\ref{fig:violation2}, we plot both the 1 and 2$\sigma$ regions for this benchmark, and we can see that the confidence region does not intersect with the positivity region (where positivity bounds are satisfied) at the 1$\sigma$ level, but does at the 2$\sigma$ level. In this case, we would conclude that the positivity region cannot be excluded at 95\% confidence-level and therefore that we have not observed positivity violation.

\begin{figure}[ht!]
	\centering
		\includegraphics[width=\textwidth, trim=20 15 55 55, clip]{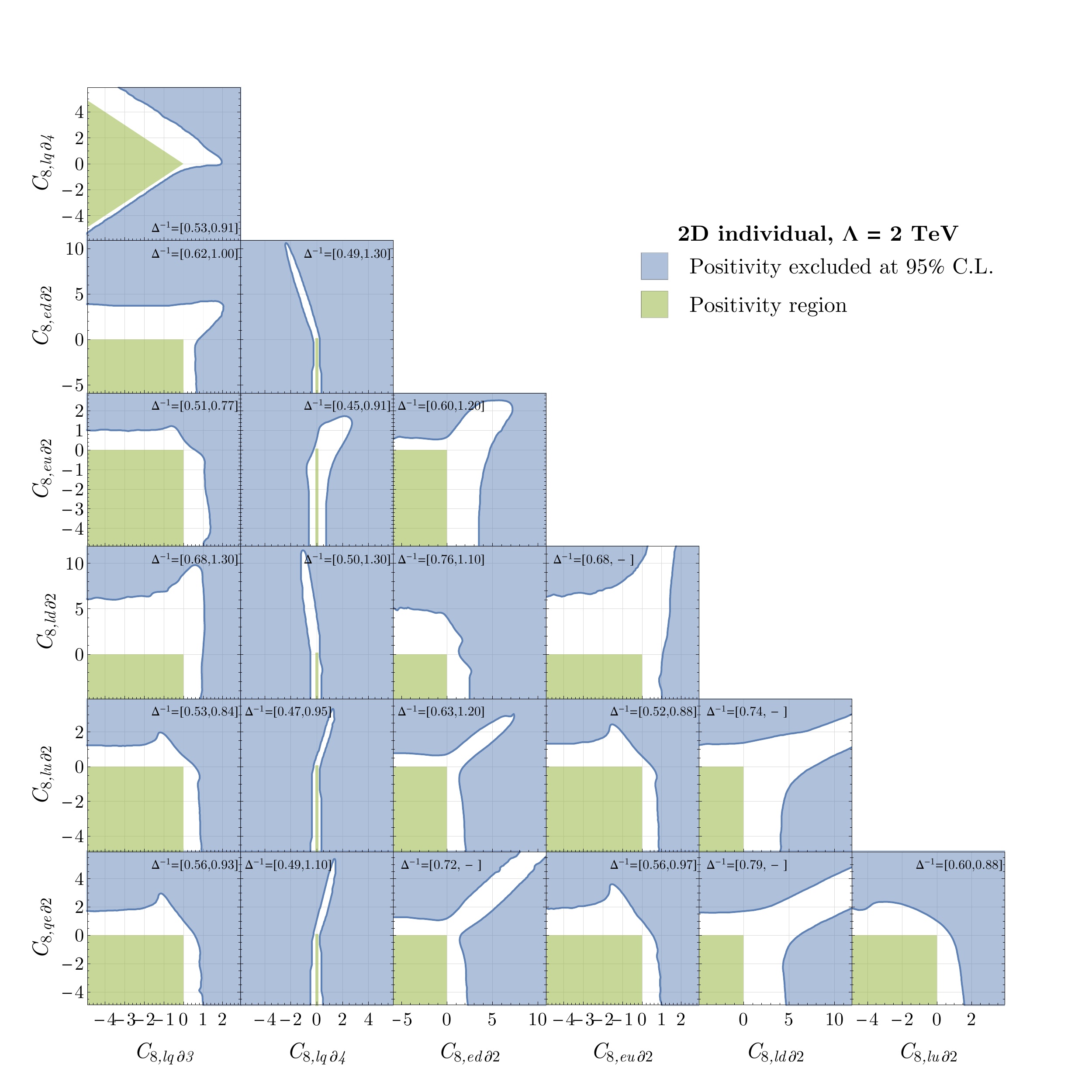}
	\caption{Sensitivity to positivity violation using the double-differential distributions of the $B_0$ and $D_0$ angular moments. The elastic positivity bounds of Table \ref{tab:elastic_bounds} are shaded in green. Blue shaded areas correspond to parameter points outside the allowed region that can be excluded at the 95 \% C.L.. We set $\Lambda= 2$~TeV in this analysis.}
	\label{fig:Pos2D}
\end{figure}
Figure~\ref{fig:Pos2D} shows the result of generalising this exercise over the 21, 2D subspaces of Wilson coefficients for $\Lambda = 2$~TeV, setting all other coefficients to zero. We shade in blue the regions in which the parameter space allowed by positivity (shaded in green) can be excluded at 95\% C.L.. This corresponds to the regions in $C_0$ where the 95\% C.L. contour does not intersect with the green areas in the figure.
We see that the sensitivity to positivity violation varies over the different 2D parameter slices, with some cases being able to probe quite close to the edge of the cone and hence larger scales. The exact ranges of $\Delta^{-1}$ probed are given in each panel. In some cases, the contour does not close around the region allowed by positivity, extending out to large positivity violation, beyond the coefficient ranges that we scanned. This highlights some remaining degeneracies in the $\chi^2$ function, that become difficult to constrain in the even of the observation of large, non-zero Wilson coefficients. The $\chi^2$ is dominated by the quadratic contributions here and the aforementioned degeneracies discussed in Table~\ref{tab:dependence} leads to some symmetries under sign flips of the coefficients, such that for a point of extreme positivity violation, there is a degeneracy with the ``mirror'' point in the extreme positivity conservation, preventing the exclusion of the allowed region. Ultimately, the exercise of distinguishing a point from the SM hypothesis (origin) as was done for Figure~\ref{fig:constrainsSM2D} is easier than that of distinguishing a point from an entire set of points in the positivity region, so it is not surprising that some additional weakly constrained directions arise. For this reason, we were not able to reliably obtain results for the profiled case.

\begin{figure}[ht!]
	\centering 
		\includegraphics[width=0.85\textwidth]{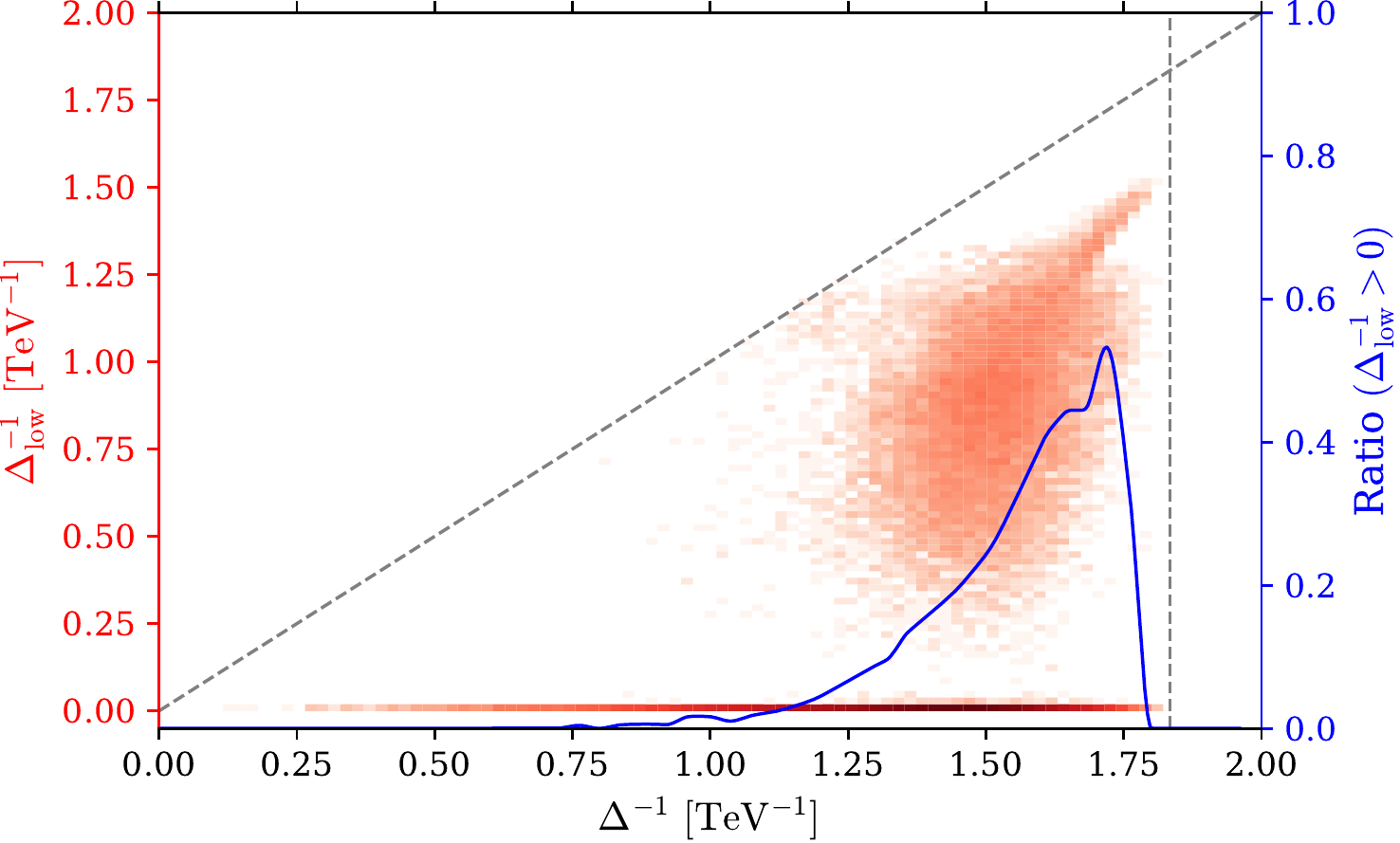}
	\caption{Distribution of positivity violation in terms of $(\Delta_{\text {low }}^{-1}, \Delta^{-1})$ at the 95 \% confidence level for the HL-LHC collider (red points, left vertical axis), for a sample of $10^5$ points in the Wilson coefficient space. We set $\Lambda= 1$~TeV in this analysis. We also plot the ratio of points whose $\Delta_{\text {low }}^{-1}$ is nonzero for a given $\Delta^{-1}$ (blue solid line, right vertical axis). $\Delta^{-1}$ does not exceed the horizontal axis value of the vertical grey dashed line that corresponds to the radius of the 7-dimensional ball.}
	\label{fig:violation1}
\end{figure}

Nevertheless, we can study the general case of allowing all seven dim-8 operators at once, by considering whether the allowed volume of parameter space intersects with the positivity region. The allowed region is bounded by the 95\% C.L. $\chi^{2}$ constraint $\chi^{2}(\vec{C}, \vec{C}_{0}) \leq \chi_{c}^{2}$, and we assume generically that the best-fit point given by the measurements can be anywhere within this region. Consequently, the $\Delta^{-1}({\vec{C}_0})$ corresponding to the best-fit point can lie within the interval:
 \begin{align}
    \Delta^{-1}({\vec{C}_0}) \in \left[ \Delta^{-1}_{\rm low} , ~ \Delta^{-1}_{\rm high} \right] ,
 \end{align}
with 
\begin{align}
    \Delta_{\text {low }}^{-1}=\min _{\chi^{2}\left(\vec{C}, \vec{C}_{0}\right) \leq \chi_{c}^{2}}\left(\frac{\delta(\vec{C}_0)^{\frac{1}{4}}}{\Lambda}\right),~~~~
    \Delta^{-1}_{\rm high} = \max _{\chi^{2}\left(\vec{C}, \vec{C}_{0}\right) \leq \chi_{c}^{2}}\left(\frac{\delta(\vec{C}_0)^{\frac{1}{4}}}{\Lambda}\right).
    \label{eq:deltalow}
\end{align}
If positivity violation can be detected, we must have $\Delta_{\text {low }}^{-1}>0$. The lower limit $\Delta_{\text {low }}$ gives a conservative estimate of the violation energy scale, {\it i.e.}, the maximum of  $\Delta$. To visualize how good the conservative estimate $\Delta_{\text {low }}^{-1}$ is compared to the ``idealized/theoretical case'' $\Delta^{-1}$ for the $\chi^{2}(\vec{C}, \vec{C}_{0})$ from the HL-LHC collider, we randomly sample the Wilson coefficient space, and plot $\Delta_{\text {low }}^{-1}$ against $\Delta^{-1}$. Figure~\ref{fig:violation1} plots the density of a random sample of $10^5$ points, uniformly distributed within a radius of $2$ from the origin of Wilso coefficient space. We set $\Lambda= 1$~TeV in this analysis.
By definition, the points in this plot have to be below the $\Delta_{\text {low }}^{-1}=\Delta^{-1}$ line. However, as this is a complex numerical minimisation problem, we cannot guarantee that the minimum obtained is the global one in every case, but only that it is a local minimum found starting from the point $\vec{C}_0$. There is an accumulation of points at the bottom of the plot with $\Delta_{\text {low }}^{-1}=0$, which is due to the fact that the $\chi^{2}(\vec{C}, \vec{C}_{0})$ region intersects with the positivity region, even though $\vec{C}_{0}$ violates positivity (see for example the 2$\sigma$ case in Figure \ref{fig:violation2}). We see that typical $\Delta_{\text {low }}^{-1}$ is roughly half of $\Delta^{-1}$, which means that the reduction in the constraining power due to the experimental uncertainties, as compared to the theoretical prediction, is usually mild for the HL-LHC. 
On the other hand, it also shows that even when $\Delta^{-1}$ is relatively large, there are still points where  $\Delta_{{\rm low}}^{-1}=0$, that is, there are still cases where we cannot unambiguously say that above a certain amount of positivity violation, we can always detect it using this set of observables. This is, again, because of the approximate flat directions and symmetries in $\chi^2$ function that we have referred to earlier. What can be ascertained is that, as the amount of positivity violation grows, we are more likely to be able to detect it. This is shown by the blue curve, which represents the fraction of points in the sample with $\Delta_{\text {low }}^{-1}>0$, \emph{i.e.}, for which a finite, upper bound on the scale of positivity violation is obtained. This can be interpreted as the probability of observing positivity violation (by ruling out the allowed positivity region).

\section{Implications for UV states}\label{sec:positivity_uv}

As mentioned previously, the convex cone/ER approach gives the strongest forward positivity bounds, and can be used to effectively infer the properties of the UV states. However, the derivation of the bounds in the general case is much more challenging, and interconnects a large number of operators. In this section, we therefore tackle a simplified problem, reduced in scope to match our purposes. As a benchmark study, we shall consider the $O_{8, eu\partial 2}$ operator alongside all possible dim-8 operators that enter into the convex cone positivity bounds involving only the right handed electron and up quark, and investigate how these can be used in conjunction with the data to reverse-engineer the UV states.

\subsection{ERs of the positivity cone and UV states}
To find the convex cone positivity bounds involving only the right handed electron and up quark, we shall focus on the following dim-8 operators~\cite{Murphy:2020rsh,Li:2020gnx}: 
\begin{align}
\label{c1op}
 C_{1}:&
\left(\bar{e} \gamma_{\mu} e\right) \partial^{2}\left(\bar{e} \gamma^{\mu} e\right)\\
 C_{2}:&
\left(\bar{e} \gamma_{\mu} e\right) \partial^{2}\left(\bar{u} \gamma^{\mu} u\right)\\
C_{3}:&(\bar{e} \gamma_{\mu} \overleftrightarrow{D}_{\nu} e ) (\bar{u} \gamma^{\mu} \overleftrightarrow{D}^{\nu} u )\\
C_{4}:& (\bar{u} \gamma_{\mu} \overleftrightarrow{D}_{\nu} u ) (\bar{u} \gamma^{\mu} \overleftrightarrow{D}^{\nu} u )\\
C_{5}:&
\left(\bar{u} \gamma_{\mu} u\right) \partial^{2}\left(\bar{u} \gamma^{\mu} u\right)
\label{c5op}
\end{align}
where $C_{8, eu\partial 2}$ has been re-named as $C_3$, and we are agnostic about all other dim-8 operators.

A (salient\footnote{A salient convex cone is one that does not contain a straight line. That is, if one half line is in a salient cone, then the opposite half line is not in the cone. The dispersion relation \eqref{ampconvexC} defines a salient cone because $m_{ij}m^*_{kl} + m_{i\tilde l}m^*_{k\tilde j}$ forms a subspace of the positive semi-definite cone, which is salient.}) convex cone can be viewed as being generated by all conical hulls (i.e., positive linear combinations) of its ERs. The most distinct property of an ER is that it cannot be conically decomposed to other rays. From dispersion relation \eqref{ampconvexC}, we see that the convex cone of the amplitude ${d^2 M_{ijkl}(0)}/{d s^2}$ can be generated by a conical hull of $m_{ij}m^*_{kl} + m_{i\tilde l}m^*_{k\tilde j}$, so the ERs must take the form of $m_{ij}m^*_{kl} + m_{i\tilde l}m^*_{k\tilde j}$. Recall that $m_{ij}$ is simply a 3-point amplitude from $i$ and $j$ to another state $X$. Therefore, an ER must have an $m_{ij}$ that cannot be split into a positive sum of other 3-point amplitudes. This is only possible if $X$ is an irreducible representation (irrep) in the product decomposition of the representations of $i$ and $j$ under the Lorentz and internal gauge symmetries of the SM. Therefore, to find the ERs, we can take a group-theoretical approach, and construct them with Clebsch-Gordan coefficients. \emph{i.e.} $m_{ij} \propto C^{r,\alpha}_{i,j}$, where $r$ runs through all irreps and $\alpha$ labels the components in the $r$ irrep. If there were only the $m_{ij}m^*_{kl}$ term in the dispersion relation, then one could obtain all ERs by finding the projectors $m_{ij}m^*_{kl}\propto P_r^{ijkl}=\sum_\alpha C^{r,\alpha}_{i,j} (C^{r,\alpha}_{k,l})^*$. Now, we actually have $m_{ij}m^*_{kl} + m_{i\tilde l}m^*_{k\tilde j}$, which means that we should take a $jl$ symmetric projection of the group projector. Under this projection, some of the ERs become non-ERs, but for many cases, especially when there are sufficient symmetries, it is relatively easy to find the real ERs among them. In short, to find the ERs of the amplitude cone, we first find potential ERs (pERs) via \cite{Zhang:2020jyn}
\begin{equation}
    (m_{ij}m^*_{kl} + m_{i\tilde l}m^*_{k\tilde j})_{\rm (pER)} \propto 
    P_r^{i(j|k|l)}=\sum_\alpha C^{r,\alpha}_{i,(j|} (C^{r,\alpha}_{k,|l)})^*
\end{equation}
and we then geometrically pick out the real ERs among them.

Alternatively, we can construct the pERs by explicitly enumerating all possible tree-level UV completions, which can be greatly aided by the group-theoretical construction  \cite{Zhang:2020jyn}, and then integrating them out to get the Wilson coefficients for the pERs. This is often an easier method in practice. Table~\ref{tab:UVstate} tabulates all of the pERs for the amplitude cone describing the positivity bounds of operators \eqref{c1op}--\eqref{c5op} and their corresponding UV particles. Our notation for the UV particles closely follows that of~\cite{deBlas:2017xtg}, and we have defined a dim-8 Wilson coefficient vector 
\begin{equation}
\label{CvecDef}
\vec{C}=(C_1,C_2,C_3,C_4,C_5)
\end{equation}
to represent the pERs. After integrating out a particular UV state $X$, we obtain a $\vec{C}$ vector for it:
\begin{equation}
\label{CXwX}
    \frac{\vec{C}_{X}}{\Lambda^4}= w_{X} \vec{c}_{X},
    ~~~w_{X}=\frac{g_{X}^{2}}{M_{X}^{4}} \geq 0
\end{equation}
where $\vec{c}_{X}$'s are the dim-8 Wilson coefficients listed in Table~\ref{tab:UVstate}, and $g_{X}$ and $M_{X}$ are generic couplings and masses of the UV states. For the partial UV states ${\cal U}_4$ and ${\cal U}_1$ in Table~\ref{tab:UVstate}, the dimensionless coupling $g_X$ should be defined as $g_X=\bar g_X M_X/M_{UV}$, where $\bar g_X/M_{UV}$ is the dimensionful coupling for the corresponding UV interactions.  In principle, there can be several copies of each UV particles of a given type, such that $w_{X}=\sum_{\cal I}{g_{X {\cal I}}^{2}}/{M_{X {\cal I}}^{4}}$ with ${\cal I}$ labeling different copies, so in Eq.~(\ref{CXwX}) we have, for simplicity, assumed that one generation dominates the summation in this paper. A generic UV completion is a combination of the dim-8 coefficients generated by several UV states
\begin{equation}
    \frac{\vec{C}}{\Lambda^4} = \sum_X \frac{\vec{C}_X}{\Lambda^4} = \sum_X w_{X} \vec{c}_{X}
\end{equation}
where we can choose to sum $X$ only over all the real ERs, since all other elements can be written as a positive sum of these. 

Two comments are in order. For our particular example with operators \eqref{c1op}--\eqref{c5op}, the tree-level one-particle UV extensions span the full space of ERs. This is not generally the case. Barring the existence of partial UV completions with higher spin particles, in some cases certain ERs can only be generated from a loop-level UV completion, in which a heavy loop diagram gives rise to a sum of the ERs that includes those that cannot be generated by tree-level one-particle UV completions. Furthermore, we have only enumerated the UV particles up to spin-1, which is sufficient for our case. We expect that it will always be the case that if the tree-level UV completions do not cover all of the ERs, adding the loop-level UV completions will. If one however restricts themselves to tree-level (partial) UV completions, some of the UV particles may have to be of higher spins, say, spin-2 massive gravitons\footnote{For a long time, it was believed that theories with massive spin-2 particles were necessarily plagued by the Boulware-Deser ghost problem~\cite{deRham:2010kj,deRham:2014zqa}. Recent studies have shown that one can construct massive spin-2 theories without the troubling Boulware-Deser ghost~\cite{deRham:2010kj, deRham:2014zqa, deRham:2016nuf}. However, these models have been conjectured to be inconsistent with the standard S-matrix axioms (which are essentially distilled from our understanding of quantum field theory without gravity), thus violating positivity bounds~\cite{Wang:2020xlt, Keltner:2015xda}. Therefore, it might be interesting to consider (potential) ERs from massive spin-2 particle, providing a way to test the fundamental principles of the S-matrix in a different sense. We leave this question for future work.}.

\begin{table}[ht]
\centering
\begin{tabular}{|c|c|c|c|}
\hline UV interaction & {\small (SU(3), SU(2))${}_{\rm U(1)}^{\rm spin}$} & dim-8 EFT coefficients ($\vec{c}_X$) & ER?\phantom{\Big|\!} \\ \hline $\bar{e^{c}} e \mathcal {S}_ 2+h.c.\huge\phantom{|}$ & $\mathcal {S}_ 2$: $(1,1)^{0}_{2}$  & $(1,0,0,0,0)$ & \CheckmarkBold \\
\hline $\frac{1}{M_{UV}} \bar{u^{c}}_{i} \overleftrightarrow{D}^{\mu} u_{j} \epsilon_{ijk}\, \mathcal{U} _ {4 \mu}^{\dagger k}+h.c.\huge\phantom{|}$ & $\mathcal{U} _ {4 \mu}^{k}$: $(\bar{3},1)^{1}_{4/3}$  & $(0,0,0,-\frac{1}{2},-\frac{3}{2})$   & \CheckmarkBold  \\
\hline $(\bar{u^{c}}_{i} u_{j}\, \Omega_ 4^{\dagger ij}+sym.)+h.c.\huge\phantom{|}$ & $\Omega _ 4^{ij}$: $(6,1)^{0}_{4/3}$  & $(0,0,0,-\frac{1}{4},\frac{1}{4})$  & \XSolidBrush \\
\hline $\bar{e^{c}} u_{i} \omega _ {1}^{\dagger i}+h.c.\huge\phantom{|}$ & $\omega _ {1}^{i}$: $(3,1)^{0}_{1/3}$  & $(0,\frac{1}{4},-\frac{1}{4},0,0)$  & \CheckmarkBold \\
\hline $\bar{e}\gamma_{\mu}u_{i}\,\mathcal {U}_{5}^{\dagger i\mu }+h.c.\huge\phantom{|}$ & $\mathcal{U}_{5}^{i\mu}$: $(3,1)^{1}_{5/3}$  & $(0,-\frac{1}{2},-\frac{1}{2},0,0)$ & \XSolidBrush \\
\hline $(\sin\theta\,\bar{e}\gamma_{\mu}e  +\cos\theta\,\bar{u}_i\gamma_{\mu}u_{i})\mathcal {B}^{\mu}\huge\phantom{|}$ & $\mathcal {B}^{\mu}$: $(1,1)^{1}_{0}$  & $(\sin^2\theta, 2\cos\theta \sin\theta ,0 , 0, \cos^2\theta)$ &\CheckmarkBold \\
\hline $\bar{u}_i\gamma_{\mu}u_{j}T_{ij}^{a}\, \mathcal {G}^{\dagger a\mu}\huge\phantom{|}$ & $\mathcal {G}^{a\mu}$: $(8,1)^{1}_{0}$  & $(0, 0, 0, -\frac {1} {4}, -\frac {5} {12})$ &  \XSolidBrush \\
\hline $\frac{1}{M_{UV}}i\bar{e^{c}} \overleftrightarrow{D}^{\mu} u^{i}\mathcal {U}_{1 \mu}^{\dagger i}+h.c.\huge\phantom{|}$ & $\mathcal {U}_{1 \mu}^{i}$: $(3,1)^{1}_{1/3}$  & $(0, -\frac {3} {4}, -\frac {1} {4}, 0, 0)$ & \CheckmarkBold \\
\hline
\end{tabular}
\caption{Potential ERs and the corresponding UV states for the amplitude cone positivity bounds of operators \eqref{c1op}--\eqref{c5op}. {\small (SU(3), SU(2))${}_{\rm U(1)}^{\rm spin}$} denotes the quantum numbers of the UV particles charged under the symmetries of the SM. The ``dim-8 EFT coefficients'' column represents pERs that can be constructed from group projectors or by integrating out the corresponding UV particles. The ``ER?'' column specifies whether a pER is really an ER or not. $e^{c}$ denotes the charge conjugate of $e$.}
\label{tab:UVstate}
\end{table}

From Table \ref{tab:UVstate}, we conclude that the pERs $\Omega_4$, $\mathcal{U}_{5}$ and $\mathcal{G}$ are not really ERs, due to the $s \leftrightarrow u$ crossing symmetric projection of $m_{ij}m^*_{kl} + m_{i\tilde l}m^*_{k\tilde j}$.
To see this, we first determine the dimensionality of the linear space spanned by all the pERs, which for our case is 5. We must then find all of the (4D in our case) hyperplanes that bound the convex cone. This is done by enumerating all of the hyperplanes spanned by any 4 of the 8 pERs and checking whether the other pERs (and thus the cone) lie on one half space delineated by that hyperplane. Having established the bounding hyperplanes (and their normal vectors), one can easily check whether or not pER is a real ER. Finally, the dot products of the normal vectors with the generic Wilson coefficient vector $\vec{C}$ give rise to the positivity bounds, \emph{i.e.}, the boundaries of the convex cone spanned by the ERs.

If all the ERs are isolated, {\it i.e.,} the cone is polytopal, all of this can be solved by a vertex/face enumeration algorithm with widely-available, efficient codes such as {\tt polymake}. However, while other ERs are isolated, the ER associated to the state $\mathcal{B}$ is parametrized by an angle $\theta$, which characterises its relative coupling strength to electrons and up-quarks.  The ER is continuous, or alternatively, represents a set of infinitely many connected ERs. This complicates the task of solving for the positivity bounds. In particular, after obtaining the positivity bounds, we also need to get rid of all continuous variables, which in our case is only the $\theta$ angle. After eliminating $\theta$, the convex cone positivity bounds for the operators \eqref{c1op}--\eqref{c5op} are given by

\begin{equation}
\begin{aligned}
\begin{split}
\label{conebound}
   & C_{3}\leq 0,\quad -3C_{4}+C_{5}\geq 0,\quad C_{4}\leq 0,\quad C_{1}\geq 0,\quad \\
   &  -(2\sqrt{C_{1}(-3C_{4}+C_{5})}-3C_{3})\leq C_2 \leq 2\sqrt{C_{1}(-3C_{4}+C_{5})}-C_{3}
\end{split}   
\end{aligned}    
\end{equation}
We will refer to the convex cone formed by these inequalities as the ${\cal C}$ cone. These (extremal/convex cone) positivity bounds on operators~\eqref{c1op}--\eqref{c5op} are obtained for the first time here. It is worth mentioning that, generally, extremal positivity bounds are stronger than elastic positivity bounds. For example, for our case, the extremal bounds on $C_3$ are stronger than the elastic scattering bound $C_3\leq 0$ (i.e., $ C_{8,eu\partial 2} \leq 0 $ in Table~\ref{tab:elastic_bounds}), when $C_4$, $C_5$, and $C_1$ satisfy the inequalities \eqref{conebound} and $C_2$ is nonzero.

\subsection{Inferring the UV states}
\begin{figure}[h]
	\centering
		\includegraphics[width=.80\linewidth]{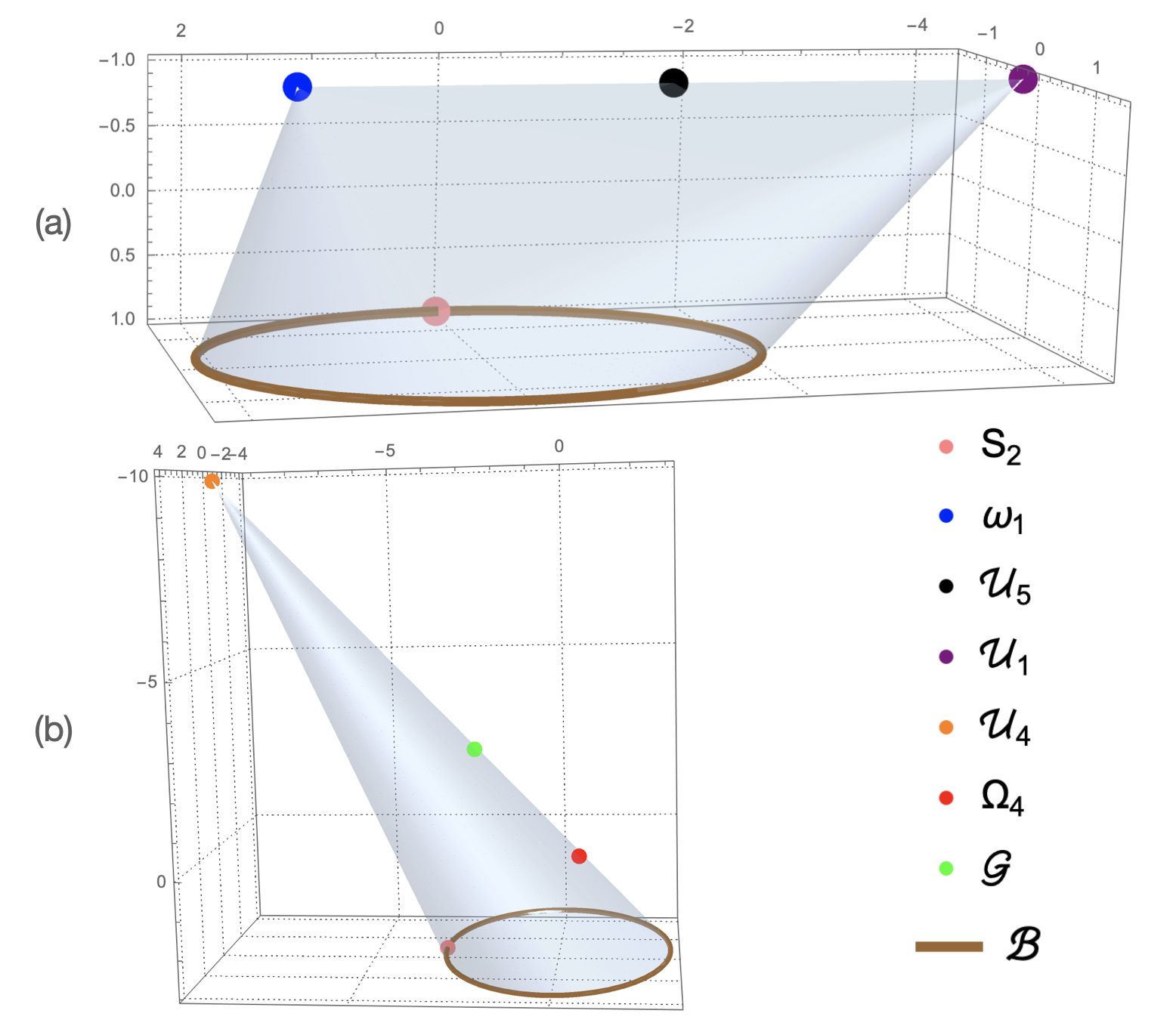}
	\caption{Two 3D cross-sections for the 5D positivity bound cone (described by (\ref{conebound})) in $eu$ scattering.
	In Subfigure (a), the $(x,y,z)$ coordinate values correspond to $(C_{2},-\tfrac{2C_{1}-3C_{3}+C_{4}+2C_{5}}{3\sqrt{2}},\tfrac{C_{5}-C_{1}}{\sqrt{2}})$, while in Subfigure  (b), the $(x,y,z)$ values correspond to $(C_{2},\frac{2C_{1}+C_{4}+2C_{5}}{3},\tfrac{C_{5}-C_{1}}{\sqrt{2}})$. 
}
  \label{fig:cone}
\end{figure}

In the previous section, we encoutered the continuous ER for the $\mathcal{B}$ state, parametrized by an angle $\theta$. This can be interpreted as the fact that the leading $s^2$-order dispersion relation cannot resolve the degeneracy between the $\mathcal{B}^\mu$ coupling to $\bar{e}\gamma_{\mu}e$ and $\bar{u}_i\gamma_{\mu}u_{i}$. The existence of continuous ERs can be easily identified in the positivity bounds~\eqref{conebound} by the presence of non-linear, square-root terms, reflecting the fact that the cone is non-polytopal. Geometrically, $\mathcal{C}$ is a 5D salient cone with six 4D faces. Since a cone consists of rays that can be infinitely extended, we can take a cross section of the 5D convex cone, resulting in a 4D convex object. To visualize this object, we take two 3D cross-sections in Figure~\ref{fig:cone}, both of which include the continuous ER. The two 3D cross-sections are representative in the sense that we are able to include all the ERs/UV states with the two cross-sections. 

In Figure \ref{fig:cone}, we see that the $\mathcal{S}_2$ ER lies on the continuous ER, which is due to the degeneracy between the $\bar{e^c} e\mathcal{S}_2$ and $\bar e \gamma_\mu e\mathcal{B}^\mu$ interaction from the viewpoint of the dispersion relation. That is, the positivity cone cannot tell the difference between ${\cal B}^\mu$ with $\theta=\pi/2$ and ${\cal S}_2$. On the other hand, we can also see from these plots that the states $\Omega_{4}$, $\mathcal{U}_{5}$ and $\mathcal{G}$ are not real ERs, as they can be decomposed as a positive sum of other states. For example, in the subfigure (a), $\mathcal{U}_5$ is not a real ER because it is in between $\mathcal{U}_1$ and $\omega_1$ and thus can be written as a positive combination of $\mathcal{U}_1$ and $\omega_1$, and in the subfigure (b), $\mathcal{G}$ and ${\Omega}_{4}$ are between $\mathcal{U}_{4}$ and $\mathcal{B}(\theta=\pi)$, so they are not real ERs.

Figure~\ref{fig:cone} presents a visualisation of how the convex cone positivity bounds can be used to infer the UV states. Assuming that the SMEFT has a standard UV completion, any future experimental measurements should point to a $\vec{C}$ vector that stays within the two geometries of Figure~\ref{fig:cone}. Of course, experimental data comes with uncertainties, and we will take this into account momentarily, but, to get the basic idea, let us for a moment assume that the data are sufficiently accurate for the energy scales we consider such that we can represent our measurements as a sharp point within the geometries. Then, for example, if the data suggest a $\vec{C}$ that is close to that of ${\cal U}_{4}$, we can predict that the UV theory must interact with the SM via a spin-1 particle that is in the $\bar 3$ representation of SU(3)$_c$, a singlet in the left handed SU(2) and with hypercharge $4/3$, when the internal symmetries of the UV theory are projected down to those of the SM. Similarly, if the future measurements find $\vec{C}$ to be at some point on the circle, we can infer that there must be a UV state $\mathcal{B}$ with (SU(3), SU(2))${}_{\rm U(1)}^{\rm spin}=(1,1)^1_0$. On the other hand, if the future collider finds $\vec{C}$ to be near $\mathcal{U}_5$, which is a pER, it may be tempting to infer that there might be a UV state with (SU(3), SU(2))${}_{\rm U(1)}^{\rm spin}=(3,1)^1_{5/3}$. However, we cannot conclude that this must be the case, as a combination of $\omega_1$ and $\mathcal{U}_{1}$ would give the same EFT coefficients. 

Since the angular dependence with $B_0$ (and $D_0$) in previous sections can only probe the $C_3\equiv C_{8, eu\partial 2}$ coefficient, it is instructive to look at the cone in the $C_3$ dimension. As we can see in Table~\ref{tab:UVstate}, only the two ERs, $\omega_1$ and ${\cal U}_{1}$, have  nonzero $C_3$ components. They both have $(3,1)_{1/3}$ as their SM gauge group quantum numbers, but differ in their Lorentz group representations, which are spin-0 and spin-1, respectively. Additionally, the pER $\mathcal{U}_5$ also has a nonzero $C_3$ component. These projected UV states are the ones that can be probed by the $B_0$ (and $D_0$) angular dependence.

To use the positivity cone to infer or exclude the UV states in practice, we also need to take into account experimental uncertainties. That is, we should combine the information on the Wilson coefficients from the positivity cone and some dataset to constrain the UV states.
First, let us assume that future experimental results are consistent with the SM. For simplicity, we shall take the resulting 95\% C.L. intervals implied by the data as the region of the parameter space in which we claim that no BSM signals are observed. That is, we shall take all $\vec{C}$ satisfying $\chi^{2}(\vec{C},\vec{C}_0)\leq \chi^{2}_{c}$ as the viable data points, where $\vec{C}_0$ is the null vector representing the observation of the SM, and $\chi_c^2$ is fixed by the desired 95\% confidence level and depends on the number of degrees of freedom. In this case, the likelihood must be constructed from data that is sensitive to the 5 Wilson coefficients for which we have constructed the positivity cone. The operators generally contribute to different scattering amplitudes and will therefore be constrained by different measurements or experiments. Our phenomenological study provides sensitivity to the $C_3$ direction, but we need to supplement the projections obtained with results from other works that have studied how to constrain the other Wilson coefficients. For instance, $C_1$ mediates $e^- e^+ \to e^- e^+$ scattering, while $C_2$ and $C_3$ could be probed by Drell-Yan (which we use in our study), $e^+e^-\to 2$  jets or Deep inelastic scattering. Finally, $C_4$ and $C_5$ mediate $u\bar{u} \to u\bar{u}$ scattering, with hadron colliders being a good candidate to probe then via, \emph{e.g.}, di-jet production. We have collected the best available constraints or projections to use in our example from the existing literature. Although this results in a rather hodgepodge collection of LHC Run 1 data and projections from HL-LHC and future electron-positron colliders, we stress that this theoretical exercise is mainly intended as an example, for which we preferred to use numbers that were at least well motivated, rather than substituting with ad hoc values. 

Coefficients that contribute to unrelated measurements will consequently not be correlated in the associated, combined likelihood. We therefore take a simplified approach, assuming that the likelihood is derived from an uncorrelated, multivariate Gaussian distribution, whose errors are characterised by the $2\sigma$ intervals taken from our work and the existing literature. Central values are taken to be $0$ in all cases, as required by the observation of the SM hypothesis.
This yields the following ``uncertainties'' for $C_1$, $C_2$ and $C_3$, for $\Lambda = 2$~TeV:
\begin{equation}
\label{C123}
C_{1}=0\pm{0.024},~~~ C_{2}=0\pm{0.45},~~~ C_{3}=0\pm{0.37}
\end{equation}
These estimates correspond individual limits or projections, neglecting all other operators, consistent with the assumption that new physics only couples to the right-handed up quark and electron\footnote{We note that these individual limits neglect the possible presence of dim-6 operators that are typically generated by the new physics scenarios we consider. Allowing for the presence of such operators would result in a somewhat weakened sensitivity that is difficult to quantify without repeating the associated studies. We expect this effect to be larger for the $C_1$ and $C_2$ bounds, since they are obtained from analyses of the full differential distributions in scattering angle and invariant mass, respectively. In our analysis of $C_3$, the impact of dim-6 operators will be reduced since they do not contribute to the $l\ge3$ moments, as discussed in Sections~\ref{sec:eft} and~\ref{sec:cutoff}.}. For $C_1$, we use the projected sensitivity obtained in~\cite{Fuks:2020ujk}, from $e^- e^+ \to e^- e^+$ scattering at the ILC running with a center-of-mass energy of $250~$GeV and with a possible upgrade to $1~$TeV (For more details, see the ILC-1000 scenario in~\cite{Fuks:2020ujk}). For $C_2$, we use the limit obtained from the process $p p \to \ell^- \ell^+$ $( \ell = e, \mu)$  at the $8$~TeV LHC with an integrated luminosity of $20.3~\mathrm{fb}^{-1}$ in table 1 of~\cite{Boughezal:2021tih}, taking the average of the upper and lower $2\sigma$ ranges. For $C_3$ we use our result, i.e., the projections for $p p \to \ell^- \ell^+$ $( \ell = e, \mu)$ at the HL-LHC  utilizing $B_0$ and $D_0$ given in Table~\ref{tab:individualbounds}, averaging over the upper and lower bounds.
To our knowledge, there are no available results for $C_4$ and $C_5$. We shall therefore consider two limiting cases: 1) $C_4=C_5=0$, 2) $C_{4},C_{5}=0\pm{10}$, where the latter case is, for our purposes, roughly equivalent to being agnostic about $C_4$ and $C_5$. Although the bounds on $C_2$ and $C_3$ come from the same underlying process (Drell Yan), we do not expect significant correlations between the two operators, since the latter bound is extracted from the $l\geq 3$ angular moments, to which the former do not contribute. Ultimately, a full, combined projection should be performed, quantifying the HL-LHC sensitivity to both operators as well as the, \emph{e.g}, di-jet sensitivity to $C_4$ and $C_5$.

Having established the allowed region from data, we now impose the positivity bounds \eqref{conebound}, which means that we require the allowed data region to be consistent with the positivity bounds (\ref{conebound}). That is, the allowed data region should have some overlap with the positivity cone ${\cal C}$. Then, we ask the following: given the experimental measurements, what is the maximum value of $w_{H}=g^2_H/M^4_H$ that is allowed for the UV state $H$ by the data and the positivity bounds? Or, equivalently, what is the lower bound, or the exclusion limit, of $M_H/\sqrt{g_H}$ for the UV state $H$ that potentially exists? This can be solved by the following optimization problem
\begin{align}
{\rm maximize}:& ~~~ \lambda\nonumber\\
{\rm subject~to}: &~~~ \vec{C}-\lambda\vec{c}_{H}\in \mathcal{C} ~~{\rm and}~~
\chi^{2}(\vec{C},\vec{C}_0)\leq \chi^{2}_{c} 
\end{align}
where for clarity $\vec{C}$ is the dim-8 Wilson coefficient vector, $\vec{c}_H$ is the ER for the UV state $H$ in Table \ref{tab:UVstate}, $\mathcal{C}$ is the positivity cone, $\vec{C}_0=\vec{0}$ and $\chi_c^2$ is determined by the 95\% confidence level. Note that $\vec{C}$ and $\lambda$ are decision variables we want to run over to maximize $\lambda$. The solution $\lambda_{\rm max}$ gives the maximum of allowed $w_{H}$, which can be converted to the lower bound of $M_H/\sqrt{g_H}$ for the UV state $H$. To see that why $\lambda_{\rm max}$ gives the maximum of $w_{H}$, we note that, as $\vec{C}$ is a positive sum of all $\vec{c}_{X}$ ERs, subtracting $\lambda\vec{c}_{H}$ from $\vec{C}$ is essentially changing $\vec{C}$ by reducing the contribution from the $\vec{c}_{H}$ ER, and the maximum of $\lambda$ is obtained when all of the $\vec{c}_{H}$ contribution is removed in the positive sum $\vec{C} /\Lambda^4 = \sum_X w_{X} \vec{c}_{X}$ and the remaining vector is still in the $\mathcal{C}$ cone. It is important to note that, thanks to our use of the positivity cone, the derived bound on the properties of a given UV state in Table~\ref{tab:UVstate} is independent of the possible existence of any of the others. The positivity cone has allowed us to make a more model-independent statement compared to the usual exercise of simply matching the SMEFT to UV models. In the latter case, the SMEFT likelihood is  typically mapped to the parameter space of a given model, assuming all of the other states are not present. In order to match the level of information used in the positivity case, one would need to interpret the data in the context of a comprehensive UV model, defined as the union of the 8 models that we consider.

\begin{table}[ht!]
\centering
\begin{tabular}{|c|c|c|}
\hline UV particle $H$  & $\lambda_{\rm max}\, [{\rm TeV}^{-4}]$ & $\frac{M_{H}}{\sqrt{g_{H}}}\, [\rm TeV])$\\
\hline $ \mathcal{S} _ 2$   & $0.0015$ & $\geq 5.1$ \\
\hline $ \mathcal{U}_{4}$ & $0$   &$\infty$  \\
\hline $\Omega_{4}$ &  0  & $\infty$ \\
\hline $\omega_{1}$ &  $0.090$  & $\geq1.8$ \\
\hline $\mathcal{U}_{5}$ &  $0.045$ &   $\geq2.2$\\
\hline $\mathcal{B}$ &0 &$\infty$ \\
\hline $\mathcal{G}$ &0 &  $\infty$ \\
\hline $\mathcal{U}_{1}$ & $0.049$ &  $\geq2.1$ \\
\hline
\end{tabular}
\caption{Lower bounds for the UV scale ${M_{H}}/{\sqrt{g_{H}}}$ assuming $C_{4}=0,C_{5}=0$. Strictly speaking, ${\cal B}$ here stands for the cases where $\theta\neq 0$; When $\theta=0$, ${\cal B}$ becomes degenerate with ${\cal S}_2$, so one cannot tell the difference between them from the view point of the positivity cone. Similar understanding applies for the later tables where $C_4=C_5=0$.}
\label{tab:UVstate1}
\end{table}
\begin{table}[ht!]
\centering
\begin{tabular}{|c|c|c|}
\hline UV particle $H$  & $\lambda_{\rm max}\, [{\rm TeV}^{-4}]$ & $\frac{M_{H}}{\sqrt{g_{H}}}\, [{\rm TeV}]$\\
\hline \hline $\mathcal{S}_2$   & 0.0015 & $\geq 5.1$ \\
\hline $\mathcal{U}_{4}$ & 1.2   &$\geq0.95$  \\
\hline $\Omega_{4}$ &  1.1  & $\geq0.97$ \\
\hline $\omega_{1}$ &  0.092  & $\geq1.8$ \\
\hline $\mathcal{U}_{5}$ &  0.046 &   $\geq2.2$\\
\hline $\mathcal{B}$ &0.00075 &$\geq6.1$ \\
\hline $\mathcal{G}$ &2.5 &  $\geq0.80$ \\
\hline $\mathcal{U}_{1}$ &  0.092 &  $\geq1.8$ \\
\hline
\end{tabular}
\caption{Lower bounds for the UV scale ${M_{H}}/{\sqrt{g_{H}}}$ assuming $C_{4}=0\pm{10},C_{5}=0\pm{10}$. This is for our purposes roughly speaking equivalent to being agnostic about $C_4$ and $C_5$.}
\label{tab:UVstate2}
\end{table}

Tables \ref{tab:UVstate1} and \ref{tab:UVstate2} summarise the results of the two cases $C_4=C_5=0$ and $C_{4}, C_{5}=0\pm{10}$ respectively. As one would expect, the case of $C_4=C_5=0$ gives stronger exclusions for the UV states. Particularly, as we see in Table \ref{tab:UVstate1} the UV states ${\cal U}_4$, $\Omega_4$, ${\cal B}$ and ${\cal G}$ are completely excluded by assuming $C_4=C_5=0$. This is not surprising as the pERs corresponding to ${\cal U}_4$, $\Omega_4$, ${\cal B}$ and ${\cal G}$ contain nonzero components for $C_4$ and $C_5$ (see Table \ref{tab:UVstate}). By a similar token, being agnostic about $C_4$ and $C_5$ in Table \ref{tab:UVstate2} only slightly weakens the lower bounds on ${\cal S}_2$, $\omega_1$, ${\cal U}_5$ and ${\cal U}_1$. However, the lower bounds on the ${\cal U}_4$, $\Omega_4$ and ${\cal G}$ state significantly relax if we do not have precise information on $C_4$ and $C_5$, while the exclusion limit on ${\cal B}$ becomes comparable with that of ${\cal S}_2$. All in all, we see that, using  our chosen dataset, we can robustly exclude the possible UV states ${\cal S}_2$, $\omega_1$, ${\cal U}_5$, ${\cal B}$ and ${\cal U}_1$ up to a few TeVs.  

\begin{table}[ht]
\centering
\begin{tabular}{|c|c|c|}
\hline UV particle $H$ & $\lambda_{\rm max}\, [{\rm TeV}^{-4}]$ & $\frac{M_{H}}{\sqrt{g_{H}}}\, [{\rm TeV}]$\\
\hline \hline $\mathcal{S}_2$   & 0.0015& $\geq 5.1$ \\
\hline $\mathcal{U}_{4}$ & 0   &$\infty$ \\
\hline $\Omega_{4}$ &  0  & $\infty$ \\
\hline $\omega_{1}$ &  0.22  & $\geq1.5$ \\
\hline $\mathcal{U}_{5}$ &  0.053 &   $\geq2.1$\\
\hline $\mathcal{B}$ &0 &$\infty$ \\
\hline $\mathcal{G}$ & 0 &  $\infty$ \\
\hline $\mathcal{U}_{1}$ & 0.053 &  $\geq2.1$ \\
\hline
\end{tabular}
\caption{Lower bounds on the UV states assuming the UV model contains only the $\omega_{1}$ particle, assuming $M_{\omega_{1}}=2$ TeV and $g_{\omega_{1}}=1$.}
\label{tab:UVstate3}
\end{table}
\begin{table}[ht]
\centering
\begin{tabular}{|c|c|c|}
\hline UV particle $H$  & $\lambda_{\rm max}\, [{\rm TeV}^{-4}]$ & $\frac{M_{H}}{\sqrt{g_{H}}}\, [{\rm TeV}]$\\
\hline \hline $\mathcal{S}_2$   & 0.0015 & $\geq 5.1$ \\
\hline $\mathcal{U}_{4}$ & 0   &$\infty$ \\
\hline $\Omega_{4}$ &  0  & $\infty$ \\
\hline $\omega_{1}$ &  0.10  & $\geq1.7$ \\
\hline $\mathcal{U}_{5}$ &  0.10 &   $\geq1.7$\\
\hline $\mathcal{B}$ &0 &$\infty$ \\
\hline $\mathcal{G}$ & 0 &  $\infty$ \\
\hline $\mathcal{U}_{1}$ & 0.17 &  $\geq1.5$ \\
\hline
\end{tabular}
\caption{Lower bounds on the UV states assuming the UV model contains only the ${\cal U}_{1}$ particle, assuming $M_{{\cal U}_{1}}=2$ TeV and $g_{{\cal U}_{1}}=1$.}
\label{tab:UVstate4}
\end{table}

On the other hand, in the event of a BSM observation from other channels in the future, we can similarly use the positivity cone to put lower bounds on the UV states, as shown in Tables~\ref{tab:UVstate3} and~\ref{tab:UVstate4}. As a crude estimate, we shall assume that the experimental uncertainties on $C_1,C_2,C_3$ are the same as those in Eq.~(\ref{C123}), even though we are now considering scenarios where the central values of $C_1,C_2,C_3$ are shifted away from zero. For a benchmark, we shall consider scenarios where the UV states have vanishing $C_4,C_5$, which allows us to limit the discussions to the subspace $C_4=C_5=0$. Under this restriction, we may assume that the BSM state is either $\omega_1$ or ${\cal U}_1$. (${\cal U}_5$ is neglected for further simplicity because it is a potential ER and cannot be uniquely determined from the dim-8 data only.) For concreteness, we further assume that the UV state has a mass of  2 TeV and the UV coupling constant $g$ is 1. Then, for the case of only the $\omega_1$ particle in the UV, we have 
$\omega_1:~C_1=0,~C_2=0.25,~C_3=-0.25$ as the central values, with $\Lambda = 2$~TeV. For the case of the ${\cal U}_1$ particle, we have ${\cal U}_1:~C_1=0,~ C_2=-0.75,~ C_3=-0.25$ as the central values. Again, the reason why the ${\cal U}_4$, $\Omega_4$, ${\cal B}$ and ${\cal G}$ states are excluded in the tables is simply because their corresponding pERs contain nonzero components for $C_4$ and $C_5$. 
We see that the lower bound for a UV state is relaxed when it is the new heavy particle, which may be taken as evidence for such a new state. (The bounds on ${\cal U}_5$ are also relaxed in both tables, particularly in Table \ref{tab:UVstate4},  because it is not a real ER but instead can be constructed from a positive sum of $\omega_1$ and ${\cal U}_1$.) Another interesting observation is that when the new state is $\omega_1$ the bounds for ${\cal U}_1$ and ${\cal U}_5$ are very close, and when the new state is ${\cal U}_1$ the bounds for $\omega_1$ and ${\cal U}_5$ are very close.

Finally, one may also wonder whether an upper bound on $M_H/\sqrt{g_H}$ can be derived for a possible UV state $H$  using the method devised in \cite{Fuks:2020ujk}, which involves constructing a new cone by flipping the sign of $H$'s ER. However, an explicit calculation shows that this is not possible if the UV state is $\omega_1$ or ${\cal U}_1$.

\section{Summary and conclusion}\label{sec:summary}
In this paper, we have explored the interplay between collider observables that are sensitive to new physics effects starting at dim-8 in the SMEFT expansion and the theoretical positivity bounds on this parameter space that arise from assuming that the associated UV completion obeys the fundamental principles of quantum field theory. Specifically we considered certain moments of the leptonic angular distribution in the Drell-Yan process at the LHC, which were recently shown to be populated by dim-8 operators.  Crucially, since these moments correspond to spherical harmonics of total angular momentum $l\geq3$, both the SM and dimension-6 operators are not able to populate them up to sub-leading EW corrections, meaning that they constitute clean probes of operators beyond dim-6.

Following a general discussion of the angular moment observables in Drell-Yan, we introduce the $B_0$ and $D_0$ moments, and discuss the relevant dim-8 operators and the reasons for which they contribute to the higher moments, while the SM and dim-6 operators do not. We then give a pedagogical review of the positivity bounds, and establish the elastic positivity bounds on our operators of interest.

We then turn to our phenomenological study, calculating the contributions from the various operators to the $B_0$ and $D_0$ moments. This is accompanied by a discussion on the information content, focused the possibility of distinguishing the effects of different operators. We observe that, since the leading, interference contributions arising at $\mathcal{O}(1/\Lambda^4)$ have the same high energy behaviour, they should not be able to be disentangled in a global fit to the Drell-Yan data. The only discriminating variable at this order is whether the operators mediate the $u\bar{u}$ or $d\bar{d}$ initial states, which motivated us to include differential information on the rapidity of the dilepton system in our analysis. We show that the contributions from the matrix element squared (of $\mathcal{O}(1/\Lambda^8)$) bring additional discriminating power: from the shape of the dilepton invariant mass distribution, the relative sign of the $B_0$ contributions at interference- and squared-order, and also by using the $D_0$ moment, which only get populated at this order. We therefore establish that the double-differential measurement in dilepton invariant mass and rapidity as containing the maximum information available to probe the $B_0$ and $D_0$ moments at LO. 

Using these results, we study the projected sensitivity of the HL-LHC dataset, assuming that we observe the SM hypothesis and that the uncertainties are statistically dominated. We find that, at the individual level, the double-differential $B_0$ moment is able to probe coefficients of order 1--2, setting $\Lambda=2$ TeV. The quadratic effects do not significantly alter the bounds, except for the two most weakly constrained operators. As expected, the rapidity information, however, is not found to have a strong impact, since it is mainly intended to be used for operator discrimination. Moving to the global case, the profiled bounds are found to be heavily dependent on using the full information. As expected, several flat directions appear at  $\mathcal{O}(1/\Lambda^4)$, which we find to completely close at  $\mathcal{O}(1/\Lambda^8)$, but not without the inclusion of $D_0$. We also study the correlations among the operators by visualising the bounds in 2D subspaces, showing how the quadratic effects and the inclusion of $B_0$ pin down the likelihood. We finally discuss how, given the sensitivity to formally higher order effects, ours results should be interpreted with care. In particular, certain dim-10 and -12 operators have not been considered, which may affect our conclusions on the sensitivity. We did, however,  intentially limit the energy range of our analysis to partly mitigate these effects.  We also point out that the dim$>$8 operators contribute to yet higher ($l>4$) moments, which our operators do not populate, meaning that the full picture would require a complete analysis of the angular distributions, also including the lower $l$ moments, which are likely to provide further information.

Next, we turn to the interesting possibility of observing the violation of positivity bounds. We therefore posit that some non-zero set of Wilson coefficients are observed and quantify the extent to which we can use the data to rule out the allowed positivity region. We show how this task is a more challenging one than simply ruling out the SM hypothesis, since one must now rule out an entire region of parameter space. 
As a basic level, this results in a slightly reduced reach in the scale of positivity violation. Looking at 2D subspaces, we also point out some approximate blind directions in our sensitivity to positivity violation, which arise due to the interplay of cancellations among $B_0$ contributions and lead to approximate symmetries in the likelihood. They prevent us from observing certain directions of large positivity violation because of our inability to exclude the mirror points, deep in the positivity allowed regions. These suggest that establishing positivity violation in a general 7D approach will be challenging using only this data. Nevertheless, we conclude the section with a probabilistic study of positivity violation in the full parameter space, finding that although it is not possible to unambiguously rule out the positivity allowed region, the probability of doing so improves with the amount of positivity violation observed.

We then examine the prospects for using our projected sensitivities as part of an exercise in combining data with the positivity cone to infer the properties of UV states. We first argue that determining the cone in the most general case is an onerous task, and permit ourselves to simplify our investigations to new physics that couples only to right-handed electrons and quarks. This corresponds to considering only one of the 7 operators we have studied alongside four new dim-8 operators that involve the same SM states. We proceed to derive and analyse the positivity cone in this space using a group-theoretical approach, connecting potential extremal rays of the cone with specific UV states that generate the operators at tree-level. The ensuing bounds have not been derived before, and represent an improvement over the elastic positivity bounds. 

Finally, we use existing and projected sensitivities from existing works alongside our projections from leptonic angular moments in  the Drell-Yan process to establish the experimental sensitivity in this space of operators. This allows us to finally quantify the ability to pin down the model space using data and the positivity cone. First assuming the observation of the SM hypothesis, the additional information provided by the cone allows us to set truly model-independent lower bounds on the scale of specific UV completions. That is, the bounds are independent of the existence of any of the other UV states that we consider. This is in contrast to the usual approach of interpreting the SMEFT likelihood directly in the context of a specific model to set lower bounds, which implicitly assumes that no other new physics is present. This allows us to set the exclusion limits on $M/\sqrt{g}$ (the ratio between the mass of the corresponding UV state and the square root of its coupling) to be a couple of TeVs. 
Assuming instead a certain benchmark model is observed, we see that the lower bounds for that particular model are significantly relaxed, while the other remain approximately unchanged, which could be used as a way to identify specific UV states in the data. 

Throughout this work, we have chosen to focus solely on the effects of dim-8 operators generated by heavy new physics. As we have discussed, this is primarily because both the observables that we consider as well as the positivity bounds are not directly sensitive to possible dim-6 operators. Furthermore it has been shown that, thanks to positivity, dim-8 operators are unavoidable in the presence of non-decoupled BSM states, while the presence of unsuppressed dim-6 effects is not guaranteed (see, e.g., \cite{Zhang:2018shp}). However, dim-6 operators are generally expected to appear, and their complete cancellation could be regarded as rather fine-tuned. However, the pattern of Wilson coefficients, and the relation to the dim-8 ones that we study are highly model-dependent. Nevertheless, we have discussed how they could contribute at higher orders to the $l\geq3$ angular moments and also how they might alter the statistical likelihood function, should non-zero values be observed in the data.

What is clear is that unsupressed dim-6 operators are likely to also leave an imprint in the data, particularly in the lower angular moments of Drell-Yan, as well as other low energy experiments. A complete picture of BSM sensitivity via the $q\bar{q}\to\ell^-\ell^+$ amplitude can only be drawn via a global analysis, incorporating all of this data. Allowing for generic, unknown dim-6 and -8 coefficients has previously been shown to significantly dilute the sensitivity of Drell-Yan, with low energy data being crucial to disentangle their effects~\cite{Boughezal:2021kla,Boughezal:2021tih}. In our case, the analysis of angular moments is expected to reduce the correlations between the two classes of operator. In the case of specific models, such as those studied in Section~\ref{sec:positivity_uv}, the combined analysis would improve the bounds on the UV states and allow, in principle, for sharper conclusions to be drawn about the model space when combining the data with positivity bounds. The question of whether the LHC data will provide sufficient sensitivity to the positivity region, given the non-observation of dimension-6 effects thus far is a poignant (albeit model-dependent) one that we leave for future investigation.

Our work is a comprehensive case study in exploiting the marriage between theoretical insights on the $S$-matrix and experimental sensitivity to a specific class of dimension-8 operators to gain unprecedented and robust sensitivity to heavy new physics beyond the SM.

\paragraph*{Acknowledgements}
KY would like to thank Kentarou Mawatari for useful discussions. The work of KY is supported by Brain Pool program funded by the Ministry of Science and ICT through the National Research Foundation of Korea (NRF-2021H1D3A2A02038697) and the Chinese Academy of Sciences (CAS) President's International Fellowship Initiative under Grant No. 2020PM0018. SYZ acknowledges support from the starting grants from University of Science and Technology of China under grant No.~KY2030000089 and GG2030040375, and is also supported by National Natural Science Foundation of China under grant No.~11947301, 12075233 and 12047502, and supported by the Fundamental Research Funds for the Central Universities under grant No.~WK2030000036. KM is supported by the UK STFC via grant ST/T000759/1.

\appendix
\section{Covariance matrix for moment functions \label{app:cov}}
In this appendix, we briefly derive the statistical covariance of a set of generic moment functions calculated over an event sample. Consider a collection of $N$ measured events corresponding to a particular integrated luminosity, $L$, over which we would like to estimate the ``un-normalized'' moment variable of a number of functions, $f_i$,
\begin{align}
    \| f_i \| \equiv\int\,d\sigma\, f_i(d\sigma) ,
\end{align}
where $d\sigma$ represents the fully differential cross section and the integrals are performed over kinematic phase space.
By ``un-normalized'', we mean that the quantities are not divided by the total cross section, in which case they would correspond to estimates of the expectation values of the $f_i(d\sigma)$. Instead, they the dimensions of a cross-section and are intended as  placeholders for the $B_0$ and $D_0$ observables considered in this work.

The total cross section, $\sigma$, and its variance, $(\delta\sigma)^2$, are estimated by
\begin{align}
\sigma = \int d\sigma \simeq \frac{N}{L},\quad (\delta\sigma)^2 = \frac{N}{L^2}.
\end{align}
To estimate the covariance matrix of $\|f_i\|$'s, we assume that the sample can be divided into a discrete set of classes with particular values of each function, $f_i^\alpha=\{f_1^\alpha,f_2^\alpha,\dots, f_n^\alpha\}$, and an associated cross section, $\sigma_a$. The value of $\| f_i \|$ is then
\begin{align}
\| f_i \| = \sum_{\alpha} \sigma_{\alpha} f_i^\alpha\simeq \frac{1}{L}\sum_{\alpha} N_{\alpha} f_i^\alpha.
\end{align}
The source of statistical uncertainty is Poissonian and is associated to the number of events, in each class, $N_{\alpha}$: $\delta N_{\alpha} = \sqrt{N_{\alpha}}$.
The estimate of the covariance of $\| f_i \|$'s, $V_{ij}$, can then be propagated, taking the $N_\alpha$'s to be uncorrelated.
\begin{align}
\begin{split}
    V_{ij} & = \sum_{\alpha} \frac{\partial \| f_i \| }{\partial N_{\alpha}}
    \frac{\partial \| f_j \| }{\partial N_{\alpha}}
    \delta N_{\alpha}^2 = 
    \frac{1}{L^2}\sum_{\alpha} 
    N_\alpha  f_i^\alpha f_j^\alpha\\
    &=\frac{1}{L}\|f_if_j \| = 
    \frac{\int \,d\sigma\,f_i(d\sigma)f_j(d\sigma)}{L}.
      \end{split}
\end{align}
Replaceing $f_i$ and $f_j$ with the correctly normalised spherical harmonic functions associated to $B_0$ and $D_0$, $Y_3^0$ and $Y_4^0$, we obtain the statistical covariance matrix in Eq.~\eqref{eq:covij}.



\end{document}